\documentclass[%
 reprint,
 amsmath,amssymb,
 aps,
]{revtex4-2}
 \usepackage{booktabs}
\usepackage{graphicx}
\usepackage{dcolumn}
\usepackage{bm}
\usepackage[table]{xcolor}
\usepackage{tikz}
\usetikzlibrary{shadings}
\usetikzlibrary{shapes, shadings}
\usepackage{amsmath}
\usepackage{amssymb}
\usepackage{slashed}

\usepackage{multirow}
\begin{document}

\preprint{APS/123-QED}

\title{A Comprehensive Search for Leptoquarks Decaying into Top-\(\tau\) Final States at the Future LHC}

\author{Debabrata Sahoo}
\email{debabrata.s@iopb.res.in}
\author{Rameswar Sahu}
\email{rameswar.s@iopb.res.in}
\author{Kirtiman Ghosh}
\email{kirti.gh@gmail.com}

\affiliation{\footnotesize Institute of Physics, Bhubaneswar, Sachivalaya Marg, Sainik School Post, Bhubaneswar 751005, India\\ \footnotesize Homi Bhabha National Institute, Training School Complex, Anushakti Nagar, Mumbai 400094, India}


\begin{abstract}
We studied the collider phenomenology of third-generation scalar leptoquarks at the Large Hadron Collider (LHC) with a 14 TeV center-of-mass energy. The analysis focuses on leptoquarks decaying exclusively into top quarks and tau leptons, employing machine learning-based tagging techniques for identifying hadronically decaying boosted top quarks, W/Z, and Higgs bosons, as well as a multivariate classifier to distinguish signal events from Standard Model (SM) backgrounds. The expected 95\% confidence level (CL) upper limits on the leptoquark production cross-section are computed assuming integrated luminosities of 200 and 500 inverse femtobarns at the 14 TeV LHC. The results demonstrate significant sensitivity improvements for detecting leptoquarks at masses beyond the current experimental limits.

\end{abstract}

\maketitle


\section{\label{sec:Intro} Introduction}

Although the Standard Model (SM), explaining strong and electroweak interaction of particles at the electroweak scale, agrees with unprecedented accuracy with the current generation of collider experiments, baring some cases like in the flavor sector, muon anomalous magnetic moment being some of the prime examples, it does not have an explanation for observed baryon asymmetry of the universe, the identity of Dark Matter and the cause of smallness of neutrino mass. Many attempts have been made to address some of these issues, which have led to various concepts and theories, such as supersymmetry, extradimensional models, left-right symmetric models, and Grand Unified Theories, to name a few.


Leptoquark is a hypothetical particle that arises naturally in many such scenarios beyond the SM (BSM) like the Pati-Salam model ~\cite{Pati:1974yy} and gauge extension to SU(5) ~\cite{Georgi:1974sy}, SO(10) ~\cite{Georgi:1974my,Fritzsch:1974nn} and other models like composite model ~\cite{Dimopoulos:1979es} and technicolor model ~\cite{Farhi:1980xs}. Leptoquarks can couple to both leptons and quarks and have color charge. Recently, there has been a boom in studies related to leptoquarks because of the interest in the flavor sector. Leptoquarks are one of the promising candidates for explaining the observed B-meson anomaly known as \(R_{K(*)}\) and \(R_{D(*)}\) \cite{LHCb:2013ghj, BaBar:2012obs, BaBar:2013mob, Fajfer:2015ycq, Barbieri:2016las, Becirevic:2015asa, Dorsner:2013tla, Bauer:2015knc, Becirevic:2016oho, Becirevic:2016yqi}, although statistical significance of \(R_{K(*)}\) is now compatible with SM prediction \cite{Belle:2019oag,LHCb:2022vje} with more data, but \(R_{D(*)}\) anomaly is still there. Also, Leptoquark is being studied in other areas, such as explaining the discrepancy in muon anomalous magnetic moment (g - 2) and neutrino mass generation ~\cite{Parashar:2022wrd}, and also, to some extent, as a portal Dark Matter ~\cite{Choi:2018stw, Mandal:2018czf}.

Leptoquarks, being charged under $SU(3)_C$, can be copiously produced at hadron collider experiments such as the Large Hadron Collider (LHC). After being pair-produced, leptoquarks decay into a quark and a SM charged lepton or neutrino, giving rise to interesting signatures at the LHC. Significant efforts have been made by the ATLAS and CMS collaborations to search for the signatures of leptoquarks in the data collected by the LHC during Run I and Run II. In the absence of any deviation of data from the SM prediction, upper limits on leptoquark production cross-sections in different decay scenarios have been imposed~\cite{ATLAS:2024fdw,ATLAS:2020dsk,CMS:2018lab,CMS:2018ncu}. Smoking gun signatures, such as an invariant mass peak in the jet-lepton invariant mass distribution resulting from leptoquarks dominantly decaying into first- or second-generation quarks and charged leptons, can be easily distinguished over the SM background. This leads to stronger bounds on the pair production of such leptoquarks. Scalar leptoquarks with masses below approximately 1.8~TeV have already been ruled out from existing LHC data~\cite{ATLAS:2020dsk,CMS:2018lab,CMS:2018ncu} for leptoquarks with 100\% branching ratios to a jet and an electron or muon. For leptoquarks with significant branching ratios into a jet and a neutrino (which remains invisible in the detector, contributing to missing energy and making the mass reconstruction of the leptoquark challenging), the limits are substantially weaker~\cite{CMS:2018lab,CMS:2018ncu}. Searches for the signatures of third-generation leptoquarks, denoted as LQ$_3^u$ and LQ$_3^d$ in the literature, which dominantly couple to third-generation quarks (top and bottom quarks) and leptons ($\tau$ and $\nu_\tau$), are challenging due to experimental limitations in reconstructing the 4-momentum of third-generation quarks and leptons, and hence the parent leptoquark masses, at the LHC. As a result, the bounds on the pair production cross-section of third-generation leptoquarks at the LHC are significantly weaker~\cite{ATLAS:2024huc,ATLAS:2021jyv,ATLAS:2023uox,ATLAS:2021oiz,ATLAS:2020dsf,ATLAS:2021yij}. For example, bounds on the third-generation leptoquarks LQ$_3^u$ and LQ$_3^d$, decaying into a third-generation quark (top and bottom quark, respectively) in association with a neutrino, are the weakest, excluding leptoquark masses below about 1.25~TeV~\cite{ATLAS:2020dsf,ATLAS:2021yij}. Whereas, for the third-generation leptoquarks LQ$_3^u$ and LQ$_3^d$ decaying into a $b\tau$ and $t\tau$ pair, respectively, the limits are relatively stronger, excluding leptoquark masses below about 1.4~TeV~\cite{ATLAS:2021jyv,ATLAS:2023uox,ATLAS:2021oiz}.

The LHC searches for third-generation leptoquarks primarily focus on conventional search strategies that rely on kinematic cuts applied to variables constructed from reconstructed physics objects such as jets, leptons, and missing transverse energy. Leptonic final states are often chosen to suppress SM background contributions. However, these final states are suppressed by the leptonic branching ratios of the $W$-boson, making this choice a trade-off between reduced background and lower signal rates. For leptoquarks with masses around 1.5~TeV decaying into a top quark, the top quark is expected to be highly boosted. In such cases, it is advantageous to reconstruct the hadronically decaying top quark as a single large-radius jet (fat jet). This approach is motivated by the enhanced hadronic decay branching ratios of the top quark compared to its leptonic decays, significantly increasing the signal rate. However, purely hadronic final states suffer from large backgrounds due to QCD multijet production. Efficient tagging of large-radius jets originating from hadronically decaying top quarks, as opposed to those from light quarks or gluons, can substantially reduce this background. Additionally, utilizing top-tagged large-radius jets allows for the kinematic reconstruction of third-generation leptoquarks decaying into top-$\tau$ pairs, further aiding in the suppression of SM backgrounds. In this work, we propose a novel search strategy for third-generation leptoquarks, specifically considering LQ$_3^d$ decaying into a top-$\tau$ pair with a 100\% branching ratio as a proof-of-concept scenario. This strategy employs a machine learning-based tagging algorithm \cite{Sahu:2024fzi,Sahu:2023uwb,Butter:2017cot} to efficiently distinguish hadronically decaying boosted top quarks, $W/Z$-bosons, and Higgs bosons from light quark and gluon jets. Additionally, multivariate classifiers are utilized for the final discrimination of the leptoquark signal over the SM background. These classifiers operate across various custom-designed signal regions, each tailored to capture different decay topologies of the leptoquarks.

The rest of the paper is structured as follows. In the Sec,~\ref{sec:model}, we have discussed the model setup and the production and decay of third-generation scalar leptoquark in our context. The phenomenology and the strategy used for the analysis are discussed in Sec.~\ref{sec:pheno}. Our results are given in Sec.~\ref{sec:result}, and finally, we conclude in Sec.~\ref{sec:conclusion}. 

\section{\label{sec:model} Model}
In this work, we are particularly interested in the collider signature of LQ$_3^d$, which dominantly decays into top quark and $\tau$-lepton pairs. While the complete list of leptoquarks along with their gauge quantum numbers under the SM gauge group, \(SU(3)_C, SU(2)_L, U(1)_Y\), can be found in Ref.~\cite{Buchmuller:1986zs}, we focus only on the scalar leptoquarks \(S_3\) $(\mathbf{\overline{3}}, \mathbf{3}, 1/3)$, \(S_1\) $(\mathbf{\overline{3}}, \mathbf{1}, 1/3)$, and \(R_2\) $(\mathbf{\overline{3}}, \mathbf{2}, 7/6)$, which can couple to our target lepton (the \(\tau\)-lepton) and quark (the $t$-quark). Here, we are particularly concerned with the third-generation scalar leptoquark that can decay into our desired final states. In addition to the usual gauge interactions of these leptoquarks with the SM gluons, $W/Z$-bosons, and photons, determined by their gauge quantum numbers, the Yukawa interactions responsible for the decay of these leptoquarks into $t\tau$ pairs are presented in the following:
\begin{align}
    \mathcal{L} &\supset y_{3\,33}^{LL} \, \bar{Q}_L^{C\, 3,a} \, \epsilon^{ab} \, (\tau^k S_3^k)^{bc} \, L_L^{3\,c} + h.c. \nonumber\\
    \mathcal{L} &\supset - y_{2\,33}^{RL} \, \bar{u}_R^{3} \, R_2^{a} \epsilon^{ab} \,  L_L^{3\,b} + y_{2\,33}^{LR} \, \bar{e}_R^{3} \, R_2^{a\, *} \,  Q_L^{3\,a} + h.c. \nonumber\\
    \mathcal{L} &\supset y_{1\,33}^{LL} \, \bar{Q}_L^{C\, 3,a} \, S_1 \epsilon^{ab} \,   L_L^{3\,b} + y_{1\,33}^{RR} \bar{u}_R^{C\, 3} S_1 e_R^j + h.c.
    \label{Yukawa}
\end{align}
While $Q_L$ and $L_L$ are the SM quark and lepton doublets, respectively, $u_R,~d_R$, and $e_R$ represent the right-handed quarks and leptons, which are singlets under the SM gauge group. The superscript 3 in the fermions denotes the third-generation fermions. The Pauli matrices are denoted by \(\tau^k\) with \(k = 1,~2,~3\), while \(a,~b\) are \(SU(2)\) indices, and \(\epsilon\) is defined as \(i\tau^2\). The superscript \(C\) represents charge conjugation, which is defined as \(\psi^C = C\bar{\psi}^T\), where \(C = i\gamma_2\gamma_0\). We assume that neutrino and quark mixing effects are negligible for our analysis, allowing us to take the PMNS~\cite{Pontecorvo:1957cp,Pontecorvo:1957qd,Pontecorvo:1967fh,Maki:1962mu} and CKM~\cite{Cabibbo:1963yz,Kobayashi:1973fv} matrices as unity. Expanding the Lagrangian in Eq.~\ref{Yukawa} and retaining only the terms relevant for the decay of the leptoquarks into a pair of top quark and $\tau$-lepton final state, we obtain:
\begin{align}
    \mathcal{L} &\supset - y_{3\,33}^{LL} \, \bar{t}_L^C {S_3^{1/3}} \tau_L + h.c. \\
    \mathcal{L} &\supset - y_{2\,33}^{RL} \, \bar{t}_R \, \tau_L R_2^{5/3} + y_{2\,33}^{LR} \, \bar{\tau}_R \,t_L R_2^{5/3 *} \,   + h.c. \\
    \mathcal{L} &\supset y_{1\,33}^{LL} \, \bar{t}_L^{C } \, S_1   \tau_L + y_{1\,33}^{RR} \bar{t}_R^{C} S_1 \tau_R + h.c.
\end{align}
Here, the superscripts on the leptoquark field represent their electric charges.

\section{Phenomenology}\label{sec:pheno}

Assuming that the Yukawa interactions in Eq.~\ref{Yukawa} are small, the dominant channel for the pair production of leptoquarks at the LHC is gluon-gluon fusion and quark-antiquark annihilation, both mediated via the strong interaction. Consequently, the leptoquark pair production cross-section depends only on the leptoquark mass and the center-of-mass energy of the proton-proton collisions at the LHC. In our analysis, we use the scalar leptoquark pair production cross-section calculated at next-to-leading order (NLO) in QCD from Ref.~\cite{Mandal:2015lca,Borschensky:2020hot}. After being produced, we assume that the scalar leptoquarks decay into the $t\tau$ final state with a branching fraction of \(100\%\). The event simulation and the reconstruction of different physics objects at the LHC with a center-of-mass energy of 14 TeV are discussed in the following sections.

We used \texttt{MadGraph5\_aMC@NLO} v3.5.5~\cite{Alwall:2014hca} with the NNPDF23 leading-order (LO) parton distribution function~\cite{Carrazza:2013axa} to generate parton-level leptoquark pair production as well as various SM background processes (discussed in detail in the next section) at the LHC with a center-of-mass energy of 14 TeV. The default renormalization and factorization scales of \texttt{MadGraph5\_aMC@NLO} have been used. The subsequent decay of unstable particles, showering, and hadronization are simulated using \texttt{PYTHIA8}~\cite{Bierlich:2022pfr}. To account for detector effects, \texttt{DELPHES}~\cite{deFavereau:2013fsa} is used with the default ATLAS card for reconstructing various physics objects such as electrons, muons, jets (\(R=0.4\) radius jets as well as fat-jets with \(R=1.2\)), missing transverse energy, etc. The details of object reconstruction are presented in the following section.

\subsection{Object Reconstruction}
For object reconstruction, we closely adhere to the methodology outlined in the ATLAS search detailed in Ref.~\cite{ATLAS:2021oiz}. Electron candidates are required to have \( p_T > 10 \, \text{GeV} \) and \( |\eta| < 2.47 \), with the additional condition that they lie outside the electromagnetic calorimeter barrel-endcap transition region. Similarly, final-state muons with \( p_T > 10 \, \text{GeV} \) and \( |\eta| < 2.5 \) are selected for our analysis. To ensure clean identification, both track-based and calorimeter tower-based isolation criteria are applied to the selected electron and muon candidates. For track-based isolation, we impose the condition \( I_R/p_T^{l} < 0.15 \), where \( I_R \) represents the sum of \( p_T \) of all tracks (excluding those of the lepton) within a cone of radius \( R_{\text{cut}} \), and \( p_T^l \) is the lepton transverse momentum. The radius \( R_{\text{cut}} \) is defined as the minimum  of 0.2 (0.3) and \( 10 \, \text{GeV}/p_T^l \) for electron (muon) candidates. For calorimeter tower-based isolation, we require that the sum of transverse energy within a cone of radius 0.2 around the lepton, excluding the lepton's energy deposit, is less than 20\% (30\%) of the electron (muon) \( p_T \).

In our analysis, we incorporate both small-radius jets (referred to as jets) and large-radius jets (commonly called fat jets). Jets (fat jets) are reconstructed using the anti-\(k_T\) algorithm~\cite{Cacciari:2008gp} with a radius parameter of 0.4 (1.2) as implemented in the \texttt{FastJet} package~\cite{Cacciari:2011ma}. The selected jets (fat jets) are required to have \( p_T > 25 \, \text{GeV} \) (\( p_T > 300 \, \text{GeV} \)) and \( |\eta| < 2.5 \). For tagging jets originating from bottom quarks and hadronically decaying $\tau$-leptons, we adopt the default algorithms and efficiencies of the ATLAS card in \texttt{DELPHES}.


\subsubsection{Boosted object tagging}

In this analysis, our goal is to exploit the presence of boosted top quarks from the decay of TeV-scale leptoquarks to suppress SM background contributions. With this motivation, we have reconstructed large-radius (\(R=1.2\)) jets, referred to as "fat jets", in the previous section. Such jets can also originate from the production of light quarks and gluons. Efficiently identifying the partonic origin of a fat jet is crucial for our analysis. We have implemented a machine learning-based algorithm to extract key fat-jet characteristics that aid in distinguishing fat jets originating from boosted top quarks, \(W/Z\), or Higgs decays from those produced by light quarks and gluons. A brief discussion on the architecture, training procedure, and classification efficiency of the classifier used in our analysis is presented in the following.

 To identify the bosted parent particle initiating the fat jets, we have used a GNN-based multiclass classifier. Given the similar masses and hadronic decay topologies of \( W \)- and \( Z \)-initiated fat jets, we group them into a single category referred to as \( V \) fat jets. For the GNN model, we closely follow the LorentzNet architecture presented in Ref.~\cite{Gong:2022lye,Sahu:2023uwb}. While the original LorentzNet architecture was developed for binary classification, one can use it for multiclass classification by updating the number of nodes in the output layer. However, such a simplified approach may not provide the optimal architecture required for the problem at hand. To find a suitable architecture, we have followed a dedicated manual hyperparameter optimization procedure. We found that by changing the number of Lorentz group equivariant blocks (LGEB) to four, the LorentzNet architecture gives the best result for the multiclass classification problem. Additionally, we have also confirmed that the design of the input block, decoding block (except for the number of nodes in the final layer), and the LGEBs work well for our problem.

For the generation of the dataset for training and testing the model, we have used a method similar to the one outlined in Ref. \cite{Gong:2022lye,Sahu:2024fzi,Sahu:2023uwb}. Top, $V$, and QCD fat jet samples are generated using the SM process $p p \rightarrow t t~$, $ p p \rightarrow W+ W-/Z Z$, $p p \rightarrow j j$, respectively. For Higgs fat jets, we used the default heft model \cite{} of MadGraph for Higgs pair production.  All other data generation steps are similar to that in Ref.~{\cite{Gong:2022lye,Sahu:2024fzi,Sahu:2023uwb}} with slight changes. Contrary to their use of mass as the node embedding scalar, we have used the charge of particles \cite{Sahu:2023uwb}. Additionally, we have generated fat jets over a wide range of transverse momentum ranging from 300 GeV to 1500 GeV. For a uniform population of fat jets in this range, we have generated 50k events for training and 25k each for testing and validating in bins of 100 GeV for each of the four categories of fat jets. 

Figure~\ref{fig:classifier-score} presents the distribution of the top ($S_T^J$), \(V\) ($S_V^J$), and \(h\) ($S_H^J$) scores in the left, middle, and right panel plots, respectively. While the classifier was trained with \(W\)- and \(Z\)-initiated jets grouped into a single class, we present the performance by treating these jets as separate categories for clarity.  In Table~\ref{tab:confusion_matrix}, we present the confusion matrix of our multiclass classifier. To obtain these results, we have used fat jets in the 500 GeV to 600 GeV transverse momentum bin. To minimize the mistagging of QCD fat jets into other categories, we adopt a straightforward threshold-based approach: any fat jet with a QCD score ($S_{QCD}^J$) greater than a predefined threshold (denoted as QCD-threshold score) is classified as a QCD fat jet. The remaining fat jets are assigned to the category with the highest score. The tagging and mistagging efficiencies in Table~\ref{tab:confusion_matrix} correspond to a QCD-threshold score of 0.4.  With this QCD-threshold, about \(90\%\) (see Table~\ref{tab:confusion_matrix}) of the truth-level tagged~\cite{Sahu:2023uwb} hadronically decaying top quark fat-jets are identified as top-tagged jets, with a rejection factor of 20 for QCD jets. A higher QCD-threshold would improve the rejection factor for QCD jets while reducing the top-tagging efficiency, which is undesirable for certain signal regions with leptons. Therefore, we do not impose a stricter QCD-threshold.  It is important to note that this second step, where fat jets are categorized into different classes, is performed solely for constructing the confusion matrix. For our phenomenological analysis, we do not assign a discrete class label to the fat jets. Instead, we retain their top/\(V\)/\(h\) scores as high-level features. These features are subsequently utilized in a boosted decision tree classifier to enhance signal and background discrimination.

\begin{widetext}
\onecolumngrid
\begin{figure}[htb!]
    \centering
    \includegraphics[width=0.32\linewidth]{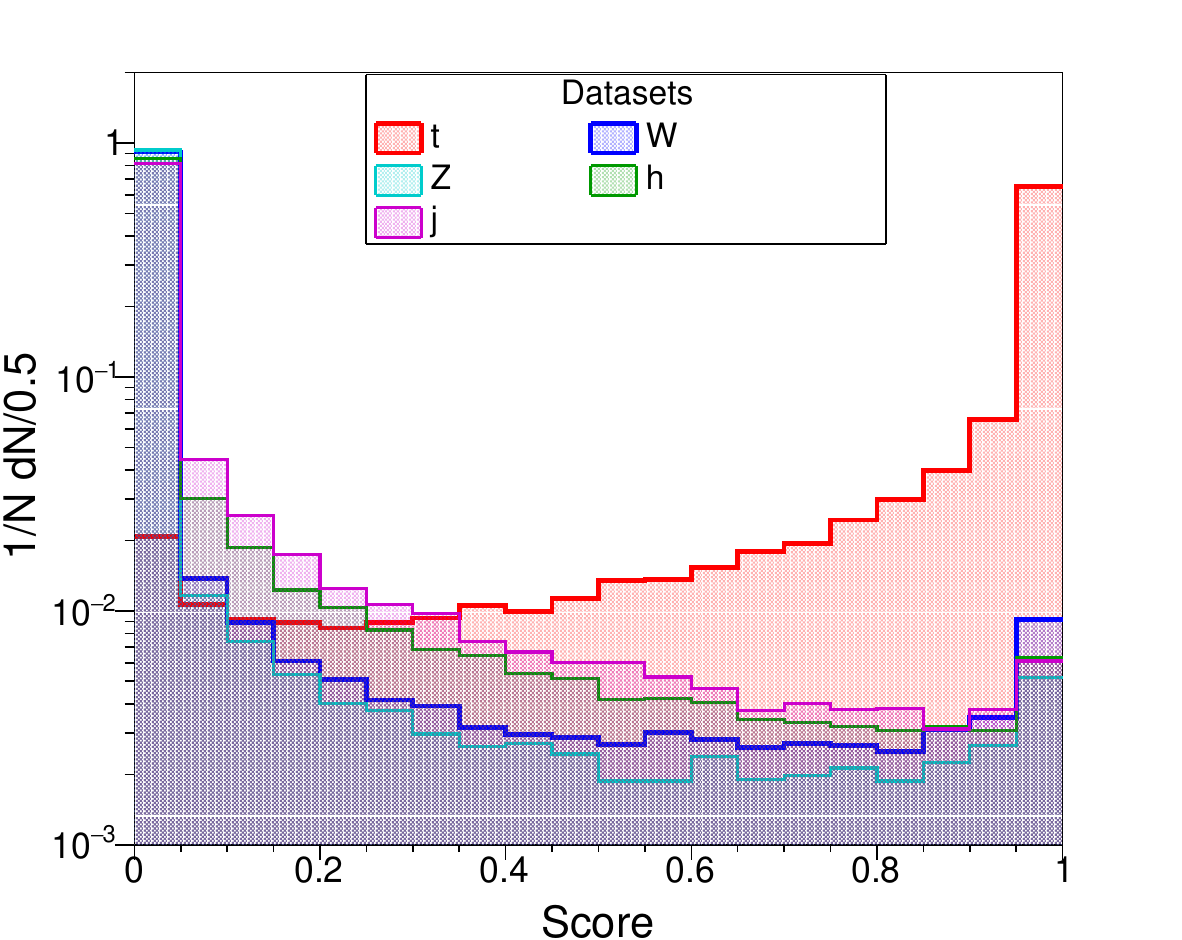}
    \includegraphics[width=0.32\linewidth]{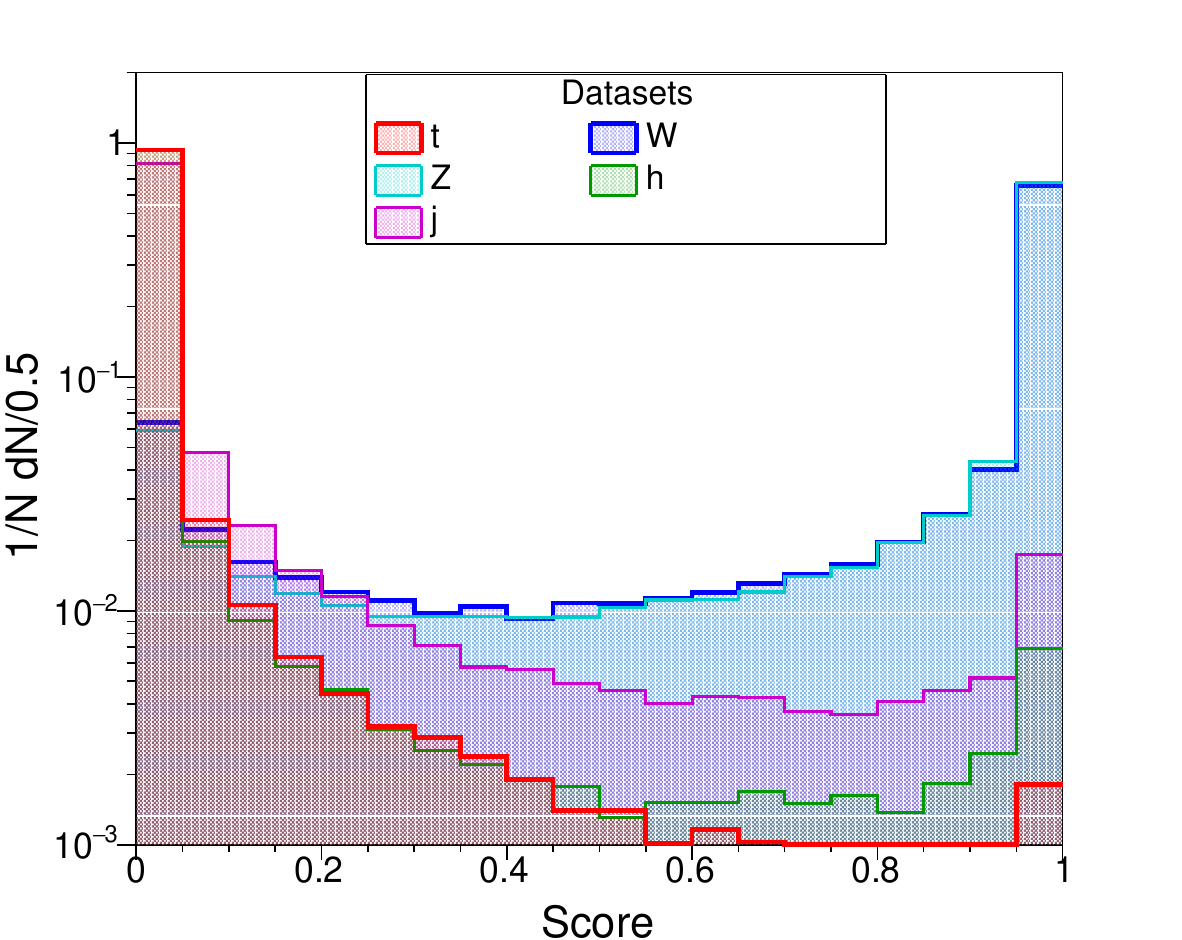}
    \includegraphics[width=0.32\linewidth]{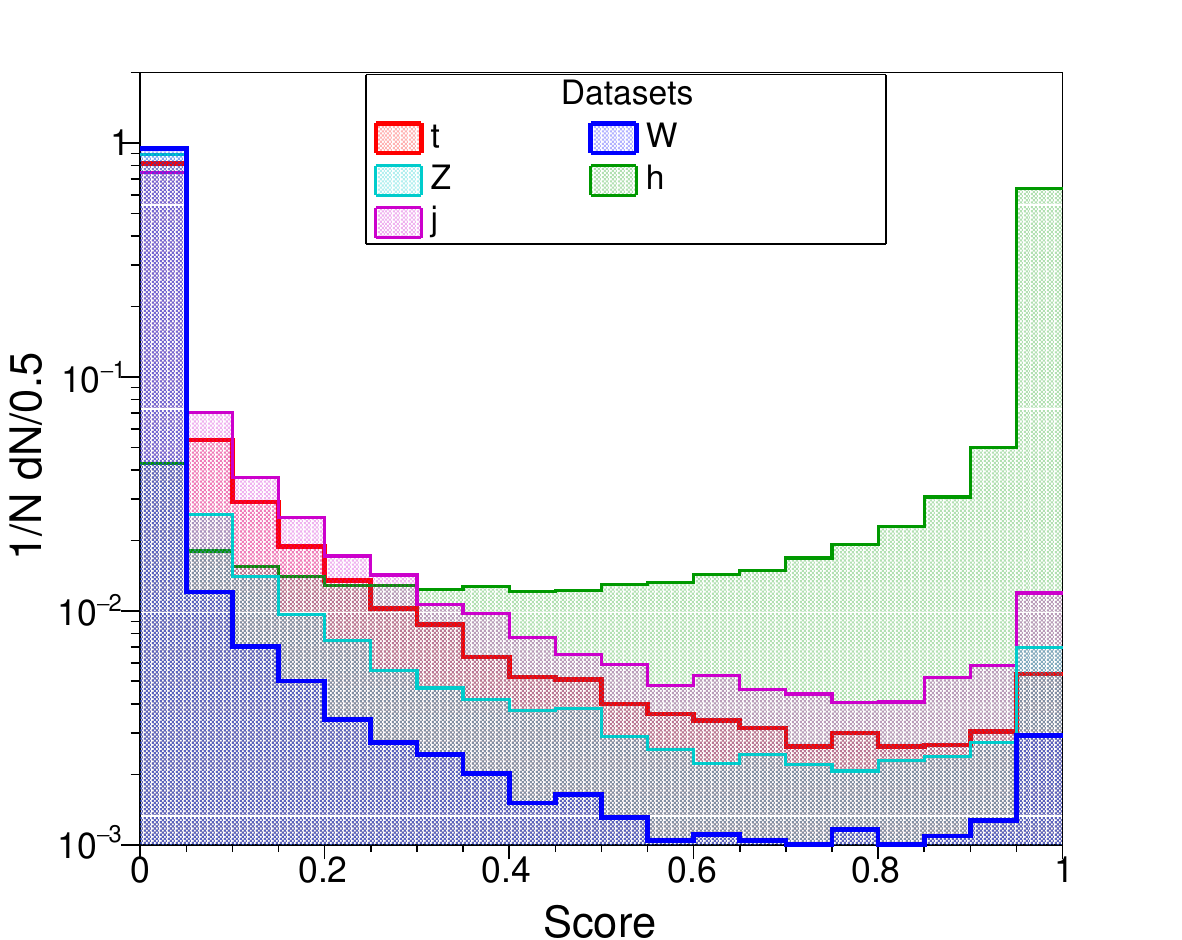}
    \caption{Normalized distribution of the top score (left), $V$ score (middle), and $h$ score (right) for the five fat jet categories.}
    \label{fig:classifier-score}
\end{figure}
\end{widetext}

\begin{table}[htb!]
    \centering
    \begin{tabular}{c|c|c|c|c}
         \toprule
         Truth label & t & v & h & j\\
         \midrule
         t & 0.902 & 0.011 & 0.038 & 0.047 \\
         W & 0.037 & 0.865 & 0.014 & 0.082 \\
         Z & 0.026 & 0.862 & 0.031 & 0.078 \\
         h & 0.043 & 0.023 & 0.859 & 0.073 \\
         j & 0.046 & 0.052 & 0.055 & 0.845 \\
         \bottomrule
         
    \end{tabular}
    \caption{Confusion matrix of the multi-class classifier with QCD score threshold of 0.4.}
    \label{tab:confusion_matrix}
\end{table}

To prevent potential double counting of selected objects, we implement the following algorithm: Electrons found within a distance of \(\Delta R = 0.1\) from a selected muon are excluded from the analysis. In cases where two electrons lie within \(\Delta R = 0.1\) of each other, only the electron with the highest \( p_T \) is retained, while the other is discarded. Additionally, small-radius jets located within a cone of \(\Delta R = 0.2\) around a selected lepton are removed to ensure a clean and unambiguous object selection.

\begin{table}[]
\begin{tabular}{|c|c|c|c|c|c|}
\hline\hline
\multicolumn{6}{|p{8.5cm}|}{\textbf{Signal Regions with lepton veto}}\\\hline \hline
 & \textbf{SRs} & $n_{J}$ & $n_{\tau_{h}}$ & $n_{b\text{-tag}}$ & \textbf{Preselection Criteria} \\ \hline\hline
\multirow{6}{*}{\rotatebox{90}{\texttt{0 lepton}}} 
&  &  &  &  & \multirow{5}{*}{\parbox{4.6cm}{$n_{j^{0.4}} \ge 4$ with $p_T^{j^{0.4}} > 300, 200, 100, 100$ GeV, \\ $n_{J^{tag}} \ge 1$, \\ $E_T^{\text{miss}} > 100$ GeV, \\ $\Delta \phi(j^{0.4}, \vec{p}_T^{\text{miss}}) > 0.2$, \\ $H_T > 1500$ GeV}} \\ 
& $0l1J1\tau_{h}$ & 1 & 1 & $\ge 2$ & \\\cline{2-5}
& $0l1J2\tau_{h}$ & 1 & $\ge 2$ & $\ge 1$ & \\ \cline{2-5}
& $0l2J1\tau_{h}$ & $\ge 2$ & 1 & $\ge 1$ & \\ \cline{2-5}
& $0l2J2\tau_{h}$ & $\ge 2$ & $\ge 2$ & $\ge 1$ & \\ 
&  &  &  &  & \\ \hline
\end{tabular}
\begin{tabular}{|c|c|c|c|c|c|c|c|c|c|}
\hline
\multicolumn{10}{|p{8.5cm}|}{\textbf{Signal Regions with at least 1 lepton}}\\\hline \hline
 & \textbf{SRs} & $n_{J}$ & $n_{\tau_h}$ & $n_{b\text{-tag}}$ & & \textbf{SRs} & $n_{J}$ & $n_{\tau_h}$ & $n_{b\text{-tag}}$ \\ \hline\hline
\multirow{6}{*}{\rotatebox{90}{\texttt{1 lepton}}} 
& $1l0J1\tau_h$ & 0 & 1 & $\ge 2$ & \multirow{6}{*}{\rotatebox{90}{\texttt{$\ge$ 2 leptons}}} & $2l0J1\tau_h$ & 0 & 1 & $\ge 1$ \\ \cline{2-5} \cline{7-10}
& $1l0J2\tau_h$ & 0 & $\ge 2$ & $\ge 1$ &  & $2l0J2\tau_h$ & 0 & $\ge 2$ & $\ge 1$ \\ \cline{2-5} \cline{7-10}
& $1l1J1\tau_h$ & 1 & 1 & $\ge 1$ & & $2l1J0\tau_h$ & 1 & 0 & $\ge 1$ \\ \cline{2-5} \cline{7-10}
& $1l1J2\tau_h$ & 1 & $\ge 2$ & $\ge 1$ & & $2l1J1\tau_h$ & 1 & 1 & $\ge 1$ \\ \cline{2-5} \cline{7-10}
& $1l2J0\tau_h$ & $\ge 2$ & 0 & $\ge 1$ & & $2l2J0\tau_h$ & $\ge 2$ & 0 & $\ge 1$ \\ \cline{2-5} \cline{7-10}
& $1l2J1\tau_h$ & $\ge 2$ & 1 & $\ge 1$ & &  &  &  & $\ge 1$\\ \hline\hline
\multicolumn{10}{|p{8.5cm}|}{\textbf{Preselection Criteria}}\\\hline \hline
\multicolumn{10}{|p{8.5cm}|}{$n_{j^{0.4}} \ge 2$, $p_T^{j^{0.4}} >$ $100, 100$ GeV, $E_T^{\text{miss}} > 100$ GeV, $\Delta \phi(j^{0.4}, \vec{p}_T^{\text{miss}}) > 0.2$} \\ \hline
\end{tabular}
\caption{Summary of signal regions and preselection criteria. The columns $n_{J}$, $n_{\tau_{h}}$ and $n_{b\text{-tag}}$ denote the number of tagged SM boosted hadronically decaying objects (top, $W$, $Z$, or Higgs), hadronically decaying tau-tagged jets, and $b$-tagged jets, respectively. $j^{0.4}$ refers to $R=0.4$ radius jets constructed using the anti-$k_T$ algorithm. $\vec{p}_T^{\text{miss}}$ represents the missing transverse momentum vector, and $H_T$ is the scalar sum of $p_T$ of all $j^{0.4}$ jets.}
\label{tab:signal_regions}
\end{table}

\subsection{Event Selection and Signal Region Definition}

After defining the various physics objects in the previous section, we are now prepared to introduce the event selection criteria and categorize the preselected events into different signal regions (SRs) for further analysis. The physics objects used to define the event preselection and SRs include $R=0.4$ radius jets (denoted by $j^{0.4}$), tagged hadronically decaying SM heavy particles reconstructed as $R=1.0$ radius jets (referred to as tagged SM boosted heavy particles and denoted by $J$), $b$-tagged and $\tau$-tagged jets, leptons (electrons and muons, denoted by $l$), and missing transverse energy (denoted by $\slashed{E}_T$).

While events with zero isolated leptons are required to satisfy the stricter preselection criteria defined in the top-left cell of Table~\ref{tab:signal_regions}, the preselection criteria for events with at least one isolated electron or muon are relatively weaker, as presented in the last row of Table~\ref{tab:signal_regions}. These stricter criteria for events with zero isolated leptons are specifically designed to suppress the significant SM contributions from QCD multi-jet production. Specifically, demanding larger $R=0.4$ jet multiplicity with harder $p_T$ thresholds, stronger requirements on $H_T$ (the scalar sum of $p_T$s of all reconstructed $R=0.4$ jets in the event), and a lower bound on the azimuthal angle between the $R=0.4$ jets and the missing transverse momentum vector, along with the missing energy cut, significantly reduce the SM QCD multi-jet contributions to the zero-lepton signal regions. Further suppression results from the requirement of the presence of at least one tagged SM-boosted heavy particle in the zero-lepton signal regions. For the signal region with two or more leptons, namely the \texttt{2 lepton} signal regions, we veto events with a same-flavour dilepton invariant mass lying within the range \( m_Z \pm 10 \) GeV to suppress the SM background contribution resulting from the production and leptonic decay of the \( Z \)-boson.

After the preselection, the events are categorized into different signal regions based on the multiplicity of isolated leptons ($n_l$), tagged SM-boosted heavy particles ($n_J$), $\tau$-tagged jets ($n_{\tau_h}$), and $b$-tagged jets ($n_b$), as depicted in Table~\ref{tab:signal_regions}. The pair production of leptoquarks, followed by their decay into a top quark and a $\tau$ lepton, results in final states containing pairs of top quarks and $\tau$ leptons. Subsequent hadronic or leptonic decays of the top quarks and $\tau$ leptons lead to various final state topologies with differing values of $n_l$, $n_J$, $n_{\tau_h}$, and $n_b$. 

The signal regions in Table~\ref{tab:signal_regions} are designed to capture these diverse final state topologies resulting from different decay cascades of the produced leptoquarks. For instance, the four \texttt{0 lepton} signal regions (see the top-left cell of Table~\ref{tab:signal_regions}) are aimed at capturing the final states arising from fully hadronic decays of the top quarks and $\tau$ leptons. At the parton level, such signal events contain a pair of $b$-quarks (from the top quark decays) and hadronically decaying $\tau$ leptons (denoted by $\tau_h$). However, the finite $b$- and $\tau_h$-tagging efficiencies of the LHC detectors significantly reduce the number of signal events if tagging is required for all $b$-jets and $\tau_h$-jets in the event. Although most hadronically decaying top quarks from the decay of heavy leptoquarks (\( m_{LQ_3^d} \geq 1.4~\text{TeV} \)) are expected to be sufficiently boosted to be fully reconstructed as fat jets with a radius of \( R=1.2 \), a significant fraction of these jets fail the final GNN-based tagging criteria and are not identified as boosted heavy SM particle jets. The different \texttt{0 lepton} signal regions demand specific numbers of tagged SM-boosted heavy particles ($J$), $b$-jets, and $\tau_h$-jets. This approach accommodates the possibility that one or more physics objects in a signal event might not be tagged due to the finite tagging efficiencies of the LHC detectors. The goal is to retain a significant fraction of signal events for further analysis, thereby enhancing the sensitivity of the LHC searches. Following the same philosophy, the \texttt{1 lepton} and \texttt{$\ge$ 2 lepton} signal regions are designed to capture the final-state topologies arising from leptoquark pair production followed by the leptonic decay of one top quark or $\tau$ lepton (\texttt{1 lepton}) and two or more top quarks or $\tau$ leptons (\texttt{$\ge$ 2 lepton}), respectively. It is important to note that all 15 signal regions defined in Table~\ref{tab:signal_regions} are statistically independent, meaning there is no overlap between the signal regions. This statistical independence ensures that the significance calculated from each signal region can be reliably combined to obtain the overall significance for the analysis.

\begin{table}
	\caption{Predicted number of signal events for a leptoquark mass of 1.5~TeV and SM background events at the LHC with a center-of-mass energy of 14~TeV and an integrated luminosity of 500~fb$^{-1}$, for different signal regions and preselection criteria defined in Table~\ref{tab:signal_regions}}
	\label{tab:signal_background_SR}
	\begin{tabular}{|l|cccccc|c|}
		\toprule
		Signal & $t\bar{t}$ & $W/Z$ & $t\bar{t}V$ & 2-4 $V$ & Multi & $4$-top & $m_{LQ_3^d}$ \\
		Region & +$tW$ & +jets & $t\bar{t}VV$ & +jets & Jet &  & 1.5TeV \\
		\midrule
		$0l1J1\tau_h$ & 8453 & 772 & 63.6 & 69.8 & $3.5\!\!\times \!\!10^{4}$ & 3.4 & 8.4 \\
		$0l1J2\tau_h$ & 573 & 30.2 & 6.7 & 5.9 & 1539 & 0.4 & 7.7 \\
		$0l2J1\tau_h$ & 5493 & 246 & 109 & 29.6 & 6620 & 10.9 & 9.0 \\
		$0l2J2\tau_h$ & 346 & 12.0 & 10.1 & 1.5 & 325 & 1.1 & 7.1 \\
		$1l0J1\tau_h$ & $1.1\!\!\times \!\!10^{5}$ & 6582 & 1268 & 783 & 17.3 & 29.8 & 2.9 \\
		$1l0J2\tau_h$ & 3345 & 95.0 & 78.9 & 15.0 & 0.5 & 2.1 & 1.0 \\
		$1l1J1\tau_h$ & $6.1\!\!\times \!\!10^{4}$ & 1428 & 1057 & 321 & 12.3 & 46.2 & 13.5 \\
		$1l1J2\tau_h$ & 1923 & 28.4 & 73.1 & 7.8 & 0.2 & 3.6 & 3.4 \\
		$1l2J0\tau_h$ & $8.2\!\!\times \!\!10^{4}$ & 1629 & 1984 & 455 & 23.7 & 172 & 7.5 \\
		$1l2J1\tau_h$ & 7895 & 99.5 & 278 & 33.3 & 1.9 & 34.4 & 11.5 \\
		$2l0J1\tau_h$ & $2.4\!\!\times \!\!10^{4}$ & 3.3 & 426 & 57.7 & 0.5 & 15.9 & 1.0 \\
		$2l0J2\tau_h$ & 386 & - & 18.9 & 1.0 & - & 0.8 & 0.1 \\
		$2l1J0\tau_h$ & $7.2\!\!\times \!\!10^{4}$ & 14.0 & 1360 & 295 & 0.9 & 117 & 5.6 \\
		$2l1J1\tau_h$ & 4094 & 0.4 & 140 & 12.8 & - & 15.7 & 2.8 \\
		$2l2J0\tau_h$ & 7697 & 0.4 & 187 & 20.8 & - & 61.1 & 3.5 \\
		\bottomrule
	\end{tabular}
\end{table}

\subsection{Discriminating the Signal over the SM Background: A Multivariate Analysis}
Our objective in this section is to isolate the signal events from the SM backgrounds for all the signal regions defined in the previous section. Discriminating the signal from the SM background involves several steps: studying signal and background events in each signal region, constructing kinematic variables to capture the differences in behavior between signal and background events, and designing constraints on these variables or combinations thereof to suppress background contributions while retaining a significant fraction of the signal events.

Determining the optimal set of kinematic cuts that sufficiently reduce the background contribution while maintaining high signal efficiency requires meticulous examination of multiple kinematic distributions for both signal and background samples. This process involves inspecting these variables individually and in various combinations. However, as the number of variables increases, this task becomes increasingly time-consuming and cumbersome. To address this challenge, a more systematic and automated approach can be achieved using machine learning-based classifiers. For our final analysis, we trained multivariate classifiers (BDT-based) for each signal region to efficiently discriminate the signal from the background. Before delving into the technical details of the multivariate classifiers—such as their hyperparameters, sample preparation, the list of kinematic variables used for training and testing, and classifier performance—we first discuss the possible SM processes that might contribute to the signal regions.

The dominant contributions to the \texttt{0 lepton} signal regions arise from top-antitop ($t\bar{t}$) pair production, followed by the hadronic decay of both top quarks. The $t\bar{t}$ production has been simulated up to one additional parton in the final state using \texttt{MadGraph} at leading order (LO) in the QCD coupling. After matching/merging using the MLM prescription \cite{Mangano:2006rw,Hoeche:2005vzu}, the cross-section is scaled with an appropriate next-to-leading order (NLO) $k$-factor obtained from Ref.~\cite{Muselli:2015kba}. Due to its large cross-section, QCD multi-jet production contributes significantly to the \texttt{0 lepton} signal regions, even after applying the preselection criteria outlined in Table~\ref{tab:signal_regions}. In our analysis, we simulate QCD 4-jet production using \texttt{MadGraph} to estimate the SM multijet contributions to different signal regions. Additionally, the production of $W/Z$ bosons in association with additional partons in the final state contributes to the \texttt{0 lepton} signal regions when the $W/Z$ bosons decay into $\tau$-lepton final states. We simulate $W/Z$-boson production in \texttt{MadGraph} with up to two additional partons in the final state at leading order (LO). The matched/merged cross-section is subsequently scaled by an appropriate next-to-leading order (NLO) $k$-factor obtained from Ref.~\cite{Catani:2009sm,Balossini:2009sa}. The \texttt{0 lepton} signal regions also receive contributions from the production of $t\bar{t}$ pairs in association with one or more SM vector bosons, such as $t\bar{t}W/Z$, $t\bar{t}WW$, $t\bar{t}ZZ$, and $t\bar{t}ZW$. Additional contributions arise from four-top-quark production, as well as di-, tri-, and tetra-boson production processes. While the di-boson production processes are simulated with up to one additional parton in the final state, the simulation of other processes does not include any additional partons. The matched/merged LO cross-sections from \texttt{MadGraph} are scaled with appropriate $k$-factors obtained from Ref.~\cite{Campbell:2011bn} to account for the effects of higher-order corrections. The dominant SM background contributions to the \texttt{1 lepton} and \texttt{2 lepton} signal regions arise from three main processes: $t\bar{t}$ production, $W/Z$+jets production, and $t\bar{t}V$ production. While the QCD multijet contributions to these signal regions are highly suppressed due to the low mis-tagging rates of jets as leptons, the contributions from the production of multiple vector bosons and 4-top quarks, though small, remain non-negligible. The estimated number of background events for 500 inverse femtoben integrated luminosity of the 14 TeV LHC resulting from different SM processes in different signal regions are summarized in   Table~\ref{tab:signal_background_SR}. The number of signal events for a leptoquark mass $m_{LQ_3^d}=1.5$ TeV are also presented in the last column of Table~\ref{tab:signal_background_SR}. The number of background events in each signal region is several orders of magnitude higher than the number of signal events (see Table~\ref{tab:signal_background_SR}). In the following, we embark on the endeavor of developing multivariate classifiers aimed at suppressing the SM backgrounds in comparison to the leptoquark signal.

\subsubsection{The Multivariate Classifiers}

To discriminate between signal events and SM background, we developed multivariate classifiers for each signal region using Boosted Decision Trees (BDTs) implemented with the \texttt{xgboost} library. For training the classifiers, simulated signal and background samples are used for each signal region. In the absence of specific knowledge about the leptoquark masses, the training signal samples consist of an equal mixture of leptoquark pair production events for leptoquark masses ranging from 1.2 to 2.0 TeV, with 100 GeV intervals. 

The training background samples are derived from the simulated events for all the SM background processes discussed in the previous section. The weightings of the signal and background events are crucial for improving the efficiency and accuracy of multivariate classifiers. The event weightings aim to maintain a balance between signal and background samples while accounting for differences in cross-section, the actual number of events present in the training dataset, etc. In our analysis, the events in the background samples from different background processes are weighted according to the production cross-section of that particular process. This ensures a realistic representation of the relative contributions of different background processes during training. Based on the number of signal and background events available for training in each signal region, the total weight of signal events is scaled as fractions (e.g., 0.2 for \texttt{0 lepton}, 0.1 for \texttt{1 lepton}, and 0.05 for \texttt{2 lepton} signal regions) of the total background weight. Similar weighting strategies have been employed in several phenomenological and experimental analyses.

\begin{table}[h]
\centering
\caption{Hyperparameters of the BDT Classifiers}
\label{tab:bdt_hyperparameters}
\begin{tabular}{|l|c|}
\hline
\textbf{Hyperparameter}            & \textbf{Value/Range}             \\ \hline
Learning Rate  & 0.1                              \\ \hline
Maximum Depth      & 4                                \\ \hline
Number of Trees  & 500                              \\ \hline
Objective Function   & \texttt{binary:logistic}        \\ \hline
Evaluation Metric & AUC     \\ \hline
Seed              & 42                              \\ \hline
Boosting Algorithm     & \texttt{gbtree}                \\ \hline
\end{tabular}
\end{table}

\begin{table}[htbp]
	\begin{tabular}{|c|l|}
		\hline
		\textbf{SR} & \textbf{Feature Importance} \\
		\hline
		\multirow{3}{*}{ \rotatebox{90}{$0l1J1\tau_h$}} & \parbox[t][1cm][c]{8cm}{\textcolor[RGB]{0,0,247}{$\slashed{E}_T$},
			\textcolor[RGB]{0,0,184}{$M_{J_1 \tau_{h_1}}$},
			\textcolor[RGB]{0,0,0}{$\Delta \phi(\slashed{E}_T, J_1)$},
			\textcolor[RGB]{0,0,0}{$\Delta \phi(\slashed{E}_T, b_1)$},
			\textcolor[RGB]{0,0,0}{$M_{\tau_{h_1} b_1}$},
			\textcolor[RGB]{0,0,0}{$S_T^{J_1}$},
			\textcolor[RGB]{0,0,0}{$M_{eff}$},
			\textcolor[RGB]{0,0,0}{s},
			\textcolor[RGB]{0,0,0}{$p_T^{\tau_{h_1}}$},
			\textcolor[RGB]{209,0,0}{$S_V^{J_1}$},
			\textcolor[RGB]{215,0,0}{$n_{Light}^{Jet}$},
			\textcolor[RGB]{222,0,0}{$\Delta \phi(\slashed{E}_T, j_2)$}} \\
		\hline
		\multirow{3}{*}{ \rotatebox{90}{$0l1J2\tau_h$}} & \parbox[t][1cm][c]{8cm}{\textcolor[RGB]{0,0,72}{$M_{J_1 \tau_{h_1}}$},
			\textcolor[RGB]{0,0,58}{$M_{J_1 \tau_{h_2}}$},
			\textcolor[RGB]{0,0,53}{$\slashed{E}_T$},
			\textcolor[RGB]{0,0,0}{$\Delta \phi(\slashed{E}_T, J_1)$},
			\textcolor[RGB]{0,0,0}{$p_T^{\tau_{h_2}}$},
			\textcolor[RGB]{0,0,0}{$\Delta \phi(\slashed{E}_T, b_1)$},
			\textcolor[RGB]{0,0,0}{$S_T^{J_1}$},
			\textcolor[RGB]{0,0,0}{$M_{\tau_{h_2} b_1}$},
			\textcolor[RGB]{0,0,0}{$\Delta \phi(\tau_{h_1}, b_1)$},
			\textcolor[RGB]{0,0,0}{$M_{eff}$},
			\textcolor[RGB]{0,0,0}{s},
			\textcolor[RGB]{215,0,0}{$n_{Light}^{Jet}$}} \\
		\hline
		\multirow{3}{*}{ \rotatebox{90}{$0l2J1\tau_h$}} & \parbox[t][1cm][c]{8cm}{\textcolor[RGB]{0,0,139}{$\slashed{E}_T$},
			\textcolor[RGB]{0,0,71}{$M_{eff}$},
			\textcolor[RGB]{0,0,28}{$M_{\tau_{h_1} b_1}$},
			\textcolor[RGB]{0,0,0}{$\Delta \phi(\slashed{E}_T, b_1)$},
			\textcolor[RGB]{0,0,0}{$M_{J_1 \tau_{h_1}}$},
			\textcolor[RGB]{0,0,0}{$M_{J_2 \tau_{h_1}}$},
			\textcolor[RGB]{0,0,0}{$S_T^{J_1}$},
			\textcolor[RGB]{0,0,0}{$\Delta \phi(\slashed{E}_T, J_2)$},
			\textcolor[RGB]{0,0,0}{$S_T^{J_2}$},
			\textcolor[RGB]{0,0,0}{s},
			\textcolor[RGB]{204,0,0}{$n_{Light}^{Jet}$},
			\textcolor[RGB]{210,0,0}{$S_V^{J_2}$}} \\
		\hline
		\multirow{3}{*}{ \rotatebox{90}{$0l2J2\tau_h$}} & \parbox[t][1cm][c]{8cm}{\textcolor[RGB]{0,0,173}{$M_{eff}$},
			\textcolor[RGB]{0,0,0}{$S_T^{J_1}$},
			\textcolor[RGB]{0,0,0}{$M_{J_2 \tau_{h_1}}$},
			\textcolor[RGB]{0,0,0}{$M_{J_1 \tau_{h_1}}$},
			\textcolor[RGB]{0,0,0}{$p_T^{\tau_{h_2}}$},
			\textcolor[RGB]{0,0,0}{$M_{\tau_{h_1} b_1}$},
			\textcolor[RGB]{0,0,0}{$\slashed{E}_T$},
			\textcolor[RGB]{0,0,0}{$M_{J_2 \tau_{h_2}}$},
			\textcolor[RGB]{0,0,0}{$\Delta \phi(\slashed{E}_T, b_1)$},
			\textcolor[RGB]{0,0,0}{$M_{\tau_{h_2} b_1}$},
			\textcolor[RGB]{0,0,0}{$n_{Light}^{Jet}$},
			\textcolor[RGB]{0,0,0}{$\Delta \phi(\slashed{E}_T, J_2)$}} \\
		\hline
		\multirow{3}{*}{ \rotatebox{90}{$1l0J1\tau_h$}} & \parbox[t][1cm][c]{8cm}{\textcolor[RGB]{0,250,0}{$M_{eff}$},
			\textcolor[RGB]{0,0,0}{$p_T^{\tau_{h_1}}$},
			\textcolor[RGB]{0,0,0}{$M_{l_1 \tau_{h_1}}$},
			\textcolor[RGB]{0,0,0}{$M_{l_1 b_1}$},
			\textcolor[RGB]{0,0,0}{$M_{\tau_{h_1} b_1}$},
			\textcolor[RGB]{0,0,0}{$M_{b_1 l_1}$},
			\textcolor[RGB]{0,0,0}{$p_T^{l_1}$},
			\textcolor[RGB]{0,0,0}{$\Delta \phi(\slashed{E}_T, b_1)$},
			\textcolor[RGB]{207,0,0}{s},
			\textcolor[RGB]{211,0,0}{$\slashed{E}_T$},
			\textcolor[RGB]{218,0,0}{$\Delta \phi(\slashed{E}_T, l_1)$},
			\textcolor[RGB]{219,0,0}{$n_{b\text{-tag}}^{l_1}$}} \\
		\hline
		\multirow{3}{*}{ \rotatebox{90}{$1l0J2\tau_h$}} & \parbox[t][1cm][c]{8cm}{\textcolor[RGB]{0,250,0}{$M_{eff}$},
			\textcolor[RGB]{0,0,56}{$p_T^{\tau_{h_1}}$},
			\textcolor[RGB]{0,0,0}{$p_T^{\tau_{h_2}}$},
			\textcolor[RGB]{0,0,0}{$M_{l_1 b_1}$},
			\textcolor[RGB]{0,0,0}{$M_{\tau_{h_1} b_1}$},
			\textcolor[RGB]{0,0,0}{s},
			\textcolor[RGB]{204,0,0}{$M_{l_1 \tau_{h_1}}$},
			\textcolor[RGB]{212,0,0}{$M_{\tau_{h_1} \tau_{h_2}}$},
			\textcolor[RGB]{213,0,0}{$\Delta \phi(\slashed{E}_T, l_1)$},
			\textcolor[RGB]{214,0,0}{$\Delta \phi(\slashed{E}_T, b_1)$},
			\textcolor[RGB]{222,0,0}{$\slashed{E}_T$},
			\textcolor[RGB]{227,0,0}{$\eta^{\tau_{h_1}}$}} \\
		\hline
		\multirow{3}{*}{ \rotatebox{90}{$1l1J1\tau_h$}} & \parbox[t][1cm][c]{8cm}{\textcolor[RGB]{0,250,0}{$M_{eff}$},
			\textcolor[RGB]{0,0,56}{$M_{J_1 \tau_{h_1}}$},
			\textcolor[RGB]{0,0,0}{$M_{J_1 l_1}$},
			\textcolor[RGB]{0,0,0}{$M_{\tau_{h_1} b_1}$},
			\textcolor[RGB]{0,0,0}{$M_{l_1 \tau_{h_1}}$},
			\textcolor[RGB]{0,0,0}{$M_{l_1 b_1}$},
			\textcolor[RGB]{0,0,0}{$\Delta \phi(\slashed{E}_T, b_1)$},
			\textcolor[RGB]{0,0,0}{$n_{b\text{-tag}}^{l_1}$},
			\textcolor[RGB]{0,0,0}{$\slashed{E}_T$},
			\textcolor[RGB]{204,0,0}{s},
			\textcolor[RGB]{215,0,0}{$\Delta \phi(\slashed{E}_T, l_1)$},
			\textcolor[RGB]{222,0,0}{$S_T^{J_1}$}} \\
		\hline
		\multirow{3}{*}{ \rotatebox{90}{$1l1J2\tau_h$}} & \parbox[t][1cm][c]{8cm}{\textcolor[RGB]{0,250,0}{$M_{eff}$},
			\textcolor[RGB]{0,0,0}{$M_{J_1 \tau_{h_1}}$},
			\textcolor[RGB]{0,0,0}{$M_{J_1 \tau_{h_2}}$},
			\textcolor[RGB]{0,0,0}{$M_{l_1 \tau_{h_2}}$},
			\textcolor[RGB]{0,0,0}{$\Delta \phi(\slashed{E}_T, b_1)$},
			\textcolor[RGB]{0,0,0}{$M_{l_1 \tau_{h_1}}$},
			\textcolor[RGB]{0,0,0}{$M_{\tau_{h_1} b_1}$},
			\textcolor[RGB]{0,0,0}{$M_{\tau_{h_1} \tau_{h_2}}$},
			\textcolor[RGB]{0,0,0}{$M_{J_1 l_1}$},
			\textcolor[RGB]{0,0,0}{$p_T^{\tau_{h_1}}$},
			\textcolor[RGB]{205,0,0}{s},
			\textcolor[RGB]{212,0,0}{$\Delta \phi(J_1, l_1)$}} \\
		\hline
		\multirow{3}{*}{ \rotatebox{90}{$1l2J0\tau_h$}} & \parbox[t][1cm][c]{8cm}{\textcolor[RGB]{0,250,0}{$M_{eff}$},
			\textcolor[RGB]{0,0,36}{$M_{l_1 b_1}$},
			\textcolor[RGB]{0,0,0}{$n_{b\text{-tag}}^{l_1}$},
			\textcolor[RGB]{0,0,0}{$M_{J_2 l_1}$},
			\textcolor[RGB]{0,0,0}{$\slashed{E}_T$},
			\textcolor[RGB]{0,0,0}{$S_T^{J_1}$},
			\textcolor[RGB]{0,0,0}{$\Delta \phi(\slashed{E}_T, J_2)$},
			\textcolor[RGB]{0,0,0}{$M_{J_1 l_1}$},
			\textcolor[RGB]{205,0,0}{$n_{b\text{-tag}}$},
			\textcolor[RGB]{212,0,0}{s},
			\textcolor[RGB]{216,0,0}{$S_V^{J_2}$},
			\textcolor[RGB]{222,0,0}{$S_T^{J_2}$}} \\
		\hline
		\multirow{3}{*}{ \rotatebox{90}{$1l2J1\tau_h$}} & \parbox[t][1cm][c]{8cm}{\textcolor[RGB]{0,250,0}{$M_{eff}$},
			\textcolor[RGB]{0,0,0}{$M_{l_1 b_1}$},
			\textcolor[RGB]{0,0,0}{$n_{b\text{-tag}}^{l_1}$},
			\textcolor[RGB]{0,0,0}{$M_{\tau_{h_1} b_1}$},
			\textcolor[RGB]{0,0,0}{$M_{J_2 \tau_{h_1}}$},
			\textcolor[RGB]{0,0,0}{$M_{J_1 \tau_{h_1}}$},
			\textcolor[RGB]{0,0,0}{$M_{J_2 l_1}$},
			\textcolor[RGB]{0,0,0}{$\slashed{E}_T$},
			\textcolor[RGB]{0,0,0}{$S_T^{J_1}$},
			\textcolor[RGB]{0,0,0}{$\Delta \phi(\slashed{E}_T, J_2)$},
			\textcolor[RGB]{0,0,0}{$M_{J_1 l_1}$},
			\textcolor[RGB]{224,0,0}{$S_T^{J_2}$}} \\
		\hline
		\multirow{3}{*}{ \rotatebox{90}{$2l0J1\tau_h$}} & \parbox[t][1cm][c]{8cm}{\textcolor[RGB]{0,250,0}{$M_{eff}$},
			\textcolor[RGB]{0,0,59}{$\sum p_T^{l}$},
			\textcolor[RGB]{0,0,0}{$M_{l_1 \tau_{h_1}}$},
			\textcolor[RGB]{0,0,0}{$M_{l_1 b_1}$},
			\textcolor[RGB]{0,0,0}{$M_{l_1 l_2}$},
			\textcolor[RGB]{0,0,0}{$\Delta \phi(\slashed{E}_T, b_1)$},
			\textcolor[RGB]{0,0,0}{$\sum Q_l$},
			\textcolor[RGB]{0,0,0}{$M_{l_2 \tau_{h_1}}$},
			\textcolor[RGB]{215,0,0}{$M_{b_1 l_1}$},
			\textcolor[RGB]{218,0,0}{$p_T^{\tau_{h_1}}$},
			\textcolor[RGB]{221,0,0}{s},
			\textcolor[RGB]{222,0,0}{$M_{l_2 b_1}$}} \\
		\hline
		\multirow{3}{*}{ \rotatebox{90}{$2l1J0\tau_h$}} & \parbox[t][1cm][c]{8cm}{\textcolor[RGB]{0,161,93}{$M_{eff}$},
			\textcolor[RGB]{0,0,254}{$M_{J_1 l_1}$},
			\textcolor[RGB]{0,0,0}{$M_{l_1 b_1}$},
			\textcolor[RGB]{0,0,0}{$n_{b\text{-tag}}^{l_1}$},
			\textcolor[RGB]{0,0,0}{$\slashed{E}_T$},
			\textcolor[RGB]{0,0,0}{$M_{l_2 b_1}$},
			\textcolor[RGB]{0,0,0}{$\Delta \phi(\slashed{E}_T, b_1)$},
			\textcolor[RGB]{0,0,0}{$S_T^{J_1}$},
			\textcolor[RGB]{0,0,0}{$\sum p_T^{l}$},
			\textcolor[RGB]{206,0,0}{$M_{J_1 l_2}$},
			\textcolor[RGB]{202,0,0}{$n_b^{l_2}$},
			\textcolor[RGB]{202,0,0}{$\sum Q_l$}} \\
		\hline
		\multirow{3}{*}{ \rotatebox{90}{$2l1J1\tau_h$}} & \parbox[t][1cm][c]{8cm}{\textcolor[RGB]{0,161,93}{$M_{eff}$},
			\textcolor[RGB]{0,0,144}{$M_{J_1 l_1}$},
			\textcolor[RGB]{0,0,0}{$M_{l_1 b_1}$},
			\textcolor[RGB]{0,0,0}{$M_{l_1 \tau_{h_1}}$},
			\textcolor[RGB]{0,0,0}{$M_{J_1 \tau_{h_1}}$},
			\textcolor[RGB]{0,0,0}{$\Delta \phi(\slashed{E}_T, b_1)$},
			\textcolor[RGB]{0,0,0}{$n_{b\text{-tag}}^{l_1}$},
			\textcolor[RGB]{204,0,0}{$\sum p_T^{l}$},
			\textcolor[RGB]{209,0,0}{$\Delta \phi(\tau_{h_1}, b_1)$},
			\textcolor[RGB]{210,0,0}{$p_T^{l_1}$},
			\textcolor[RGB]{212,0,0}{$S_T^{J_1}$},
			\textcolor[RGB]{214,0,0}{$\sum Q_l$}} \\
		\hline
		\multirow{3}{*}{ \rotatebox{90}{$2l2J0\tau_h$}} & \parbox[t][1cm][c]{8cm}{\textcolor[RGB]{0,161,93}{$M_{eff}$},
			\textcolor[RGB]{0,0,0}{$n_{b\text{-tag}}^{l_1}$},
			\textcolor[RGB]{0,0,0}{$M_{l_2 b_1}$},
			\textcolor[RGB]{0,0,0}{$M_{J_2 l_1}$},
			\textcolor[RGB]{0,0,0}{$M_{l_1 b_1}$},
			\textcolor[RGB]{0,0,0}{$S_T^{J_1}$},
			\textcolor[RGB]{0,0,0}{$M_{J_1 l_1}$},
			\textcolor[RGB]{0,0,0}{$\slashed{E}_T$},
			\textcolor[RGB]{0,0,0}{$\Delta \phi(\slashed{E}_T, J_2)$},
			\textcolor[RGB]{0,0,0}{$\sum p_T^{l}$},
			\textcolor[RGB]{0,0,0}{$n_b^{l_2}$},
			\textcolor[RGB]{0,0,0}{$\Delta \phi(\slashed{E}_T, b_1)$}} \\
		\hline
		
		\multicolumn{2}{|c|}{\includegraphics[width=8cm]{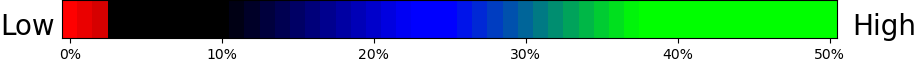}} \\
		\hline
	\end{tabular}
	\caption{List of the 12 most important kinematic variables for each signal region, as defined in Table~\ref{tab:signal_regions}, used in the BDT analysis for discriminating the signal from the SM backgrounds. The relative importance of these variables is indicated by the color coding. The symbols used for different kinematic variables are defined in the text.}
	\label{tab:feature_importance}
\end{table}

The input features for the BDTs are carefully constructed kinematic variables. These include two broad classes of features:\\
\noindent {\bf Universal Kinematic Features (UKFs)}: Universal Kinematic Features are kinematic variables that are universally calculable across all signal regions. These include: Missing transverse energy (\(\slashed{E}_T\)), \(H_T\) defined as \(H_T = \sum_{j} p_T^j\), effective mass (\(M_{eff}\)) defined as the scalar sum of \(H_T\), \(\slashed{E}_T\), and the transverse momenta of all isolated leptons in the event, Sphericity defined as $S = \frac{3}{2} \frac{\lambda_2 + \lambda_3}{\lambda_1 + \lambda_2 + \lambda_3}$  where \(\lambda_1 \geq \lambda_2 \geq \lambda_3\) are the eigenvalues of the momentum tensor: $S_{ij} = \sum_k \frac{p_{k,i} p_{k,j}}{\sum_k |\vec{p}_k|^2}$ and \(p_{k,i}\) represents the \(i\)-th component of the momentum of the \(k\)-th $R=0.4$ jet in the event, etc. Note that all the signal regions contain at least two $R=0.4$ radius jets ($j^{0.4}$) and at least one $b$-tagged jet. Several universal kinematic features are constructed using the $R=0.4$ radius jets and the leading $b$-tagged jet. These include the number of $b$-tagged jets, denoted as $n_{b\text{-tag}}$, and the number of $R=0.4$ radius jets that are not tagged as $b/\tau$-jets. These untagged jets are referred to as light jets ($l\text{-jets}$), and their count is denoted by $n_{\text{Light}}^{\text{Jet}}$. Another key feature is the scalar sum of the transverse momenta of $b$-tagged jets, represented as $\sum p_T^{b\text{-tag}}$. The transverse momentum ($p_T^{b_1}$) and rapidity ($\eta^{b_1}$) of the leading $b$-tagged jet are also used, along with the azimuthal separation between the leading $b$-tagged jet and the missing transverse energy vector, expressed as $\Delta \phi(\slashed{E_T}, b_1)$. Furthermore, the number of light jets within a radius of $\Delta R = 1.2$ around the leading $b$-tagged jet, denoted by $n^{b_1}_{l\text{-jet}}$, and the invariant mass of all the $R=0.4$ radius jets within the same $\Delta R$ radius around the leading $b$-tagged jet, denoted by $M^{b_1}_{\Delta R_{1.2}}$, are also included in the list of UKFs.\\

\noindent {\bf Signal Region-Specific Features (SRSFs)}: The Signal Region-Specific Features (SRSFs) include various kinematic variables for all the reconstructed objects (denoted as SR-objects) such as the $R=1.2$ radius jets, $\tau$-tagged jets, and leptons, which define the signal regions. The list of SRSFs comprises transverse momentum ($p_T^\text{obj}$), rapidity ($\eta^\text{obj}$), azimuthal separation with the missing transverse energy vector ($\Delta \phi(\slashed{E_T}, \text{obj})$), all possible combinations of invariant mass ($M_{\text{obj}_i\text{obj}_j}$), and azimuthal separation ($\Delta \phi(\text{obj}_i, \text{obj}_j)$) between the SR-objects. For signal regions involving one or more tagged SM-boosted heavy particles (denoted as $J$), additional features include the top score ($S_T^J$), V score ($S_V^J$), and Higgs score ($S_H^J$) for each SR-$J$. These scores are derived from a GNN-based multiclass classifier used to identify the particle origin of the $R=1.2$ radius jets and are particularly effective at reducing QCD background contributions. Furthermore, kinematic variables such as the number of $b$-tagged jets within a radius of $\Delta R=1.2$ around the tagged SM-boosted heavy objects ($n_{b\text{-tag}}^J$) and around leptons ($n_{b\text{-tag}}^l$) are included. These variables are instrumental in identifying boosted top quarks decaying hadronically and leptonically, respectively.

\begin{widetext}
\onecolumngrid

\begin{figure}[h]
    \centering
    \includegraphics[width=0.32\textwidth]{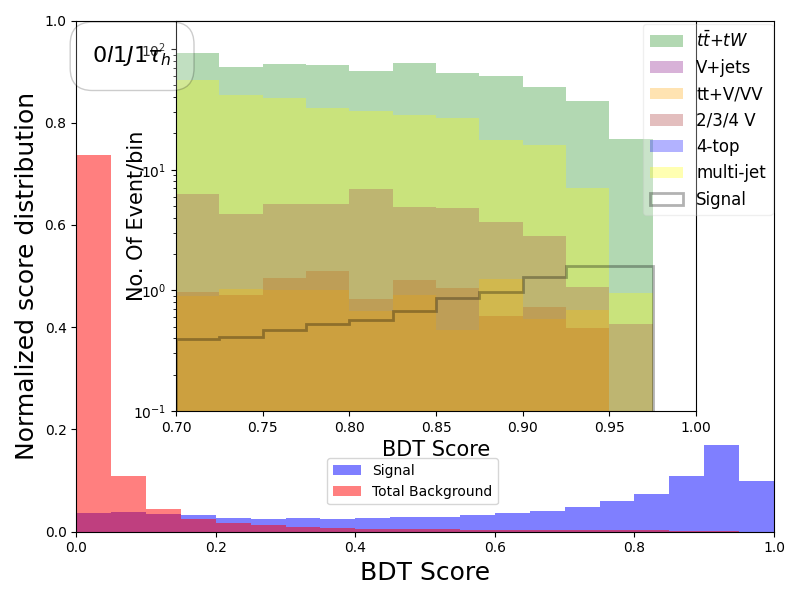}
    \includegraphics[width=0.32\textwidth]{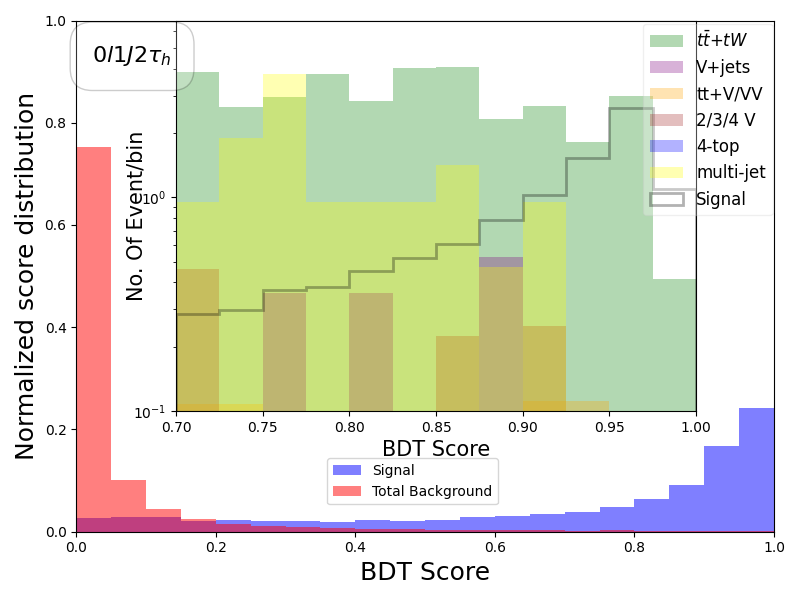}
    \includegraphics[width=0.32\textwidth]{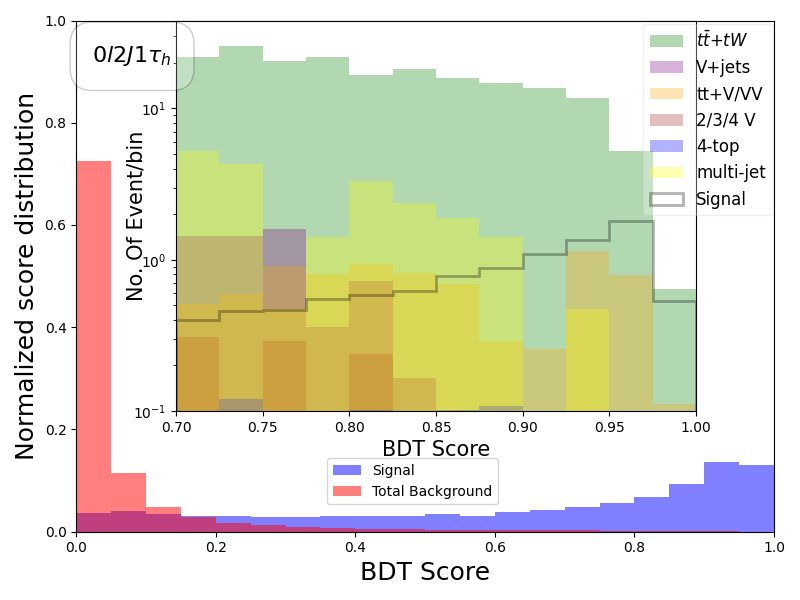}\\
    \includegraphics[width=0.32\textwidth]{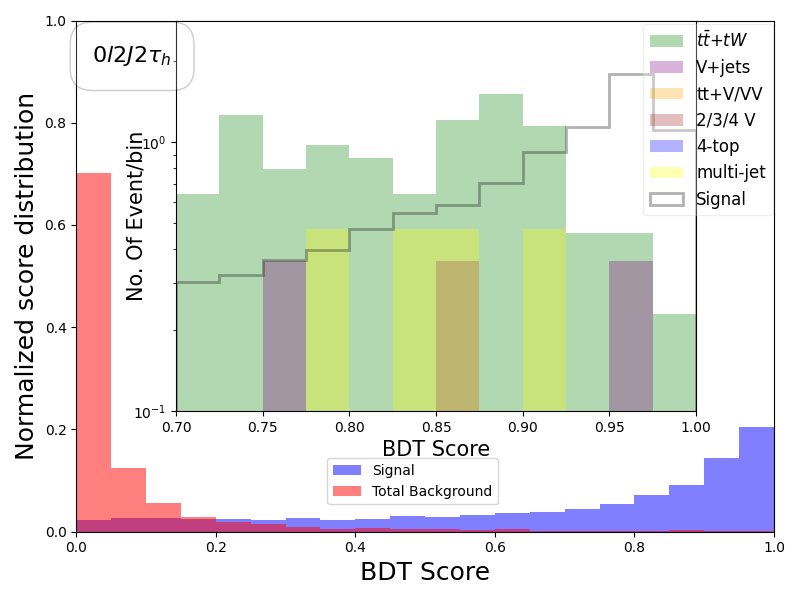}
    \includegraphics[width=0.32\textwidth]{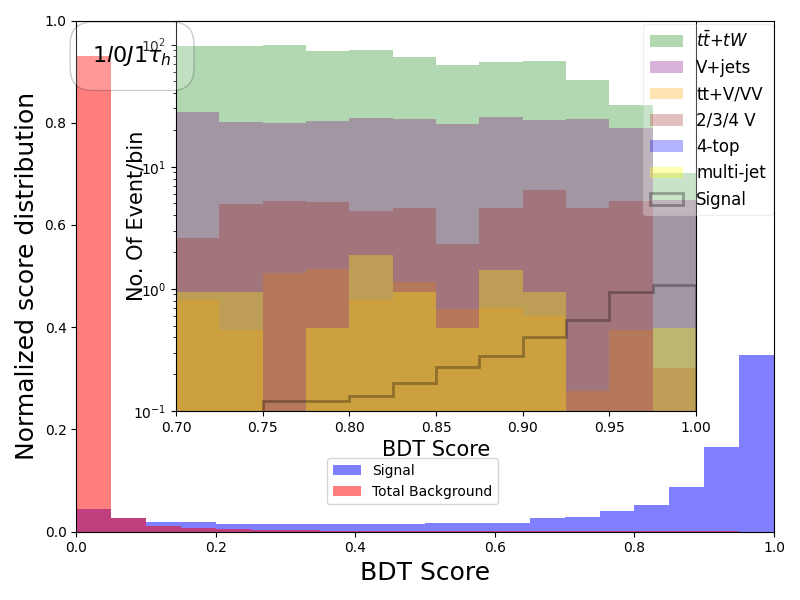}
    \includegraphics[width=0.32\textwidth]{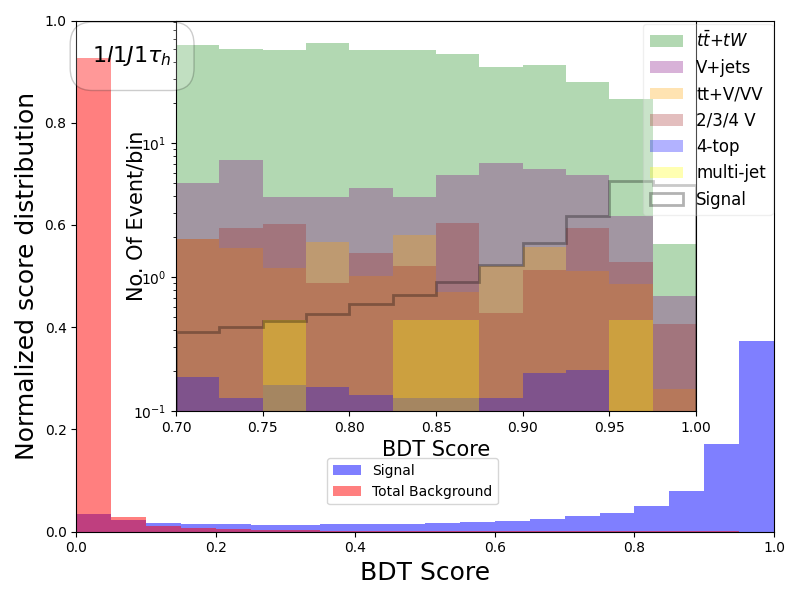}\\
    \includegraphics[width=0.32\textwidth]{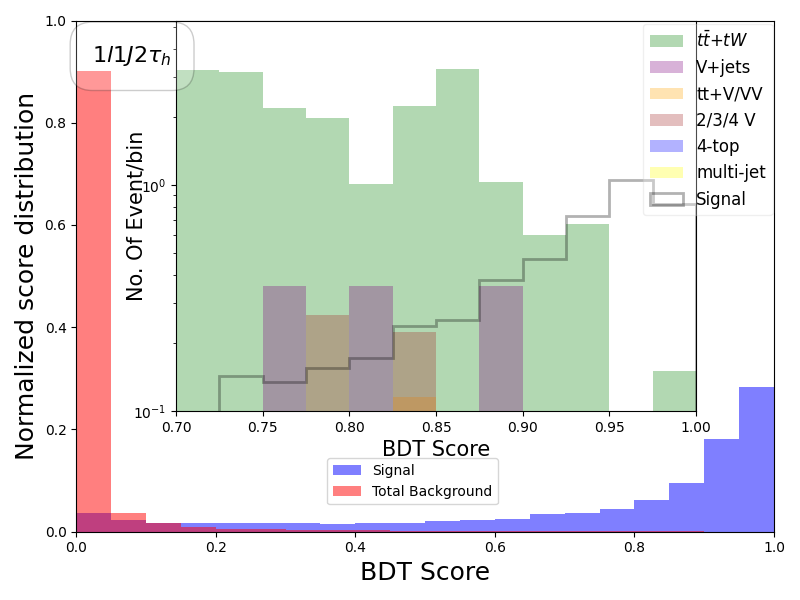}
    \includegraphics[width=0.32\textwidth]{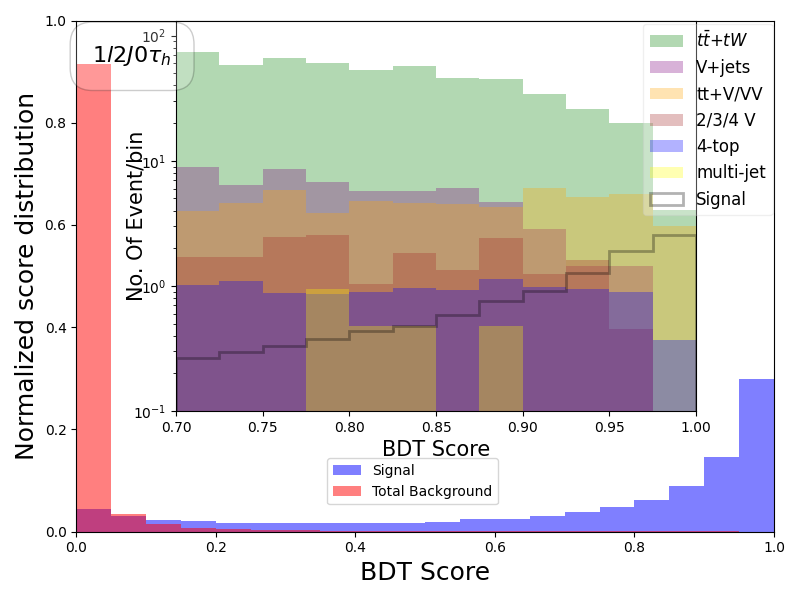}     
    \includegraphics[width=0.32\textwidth]{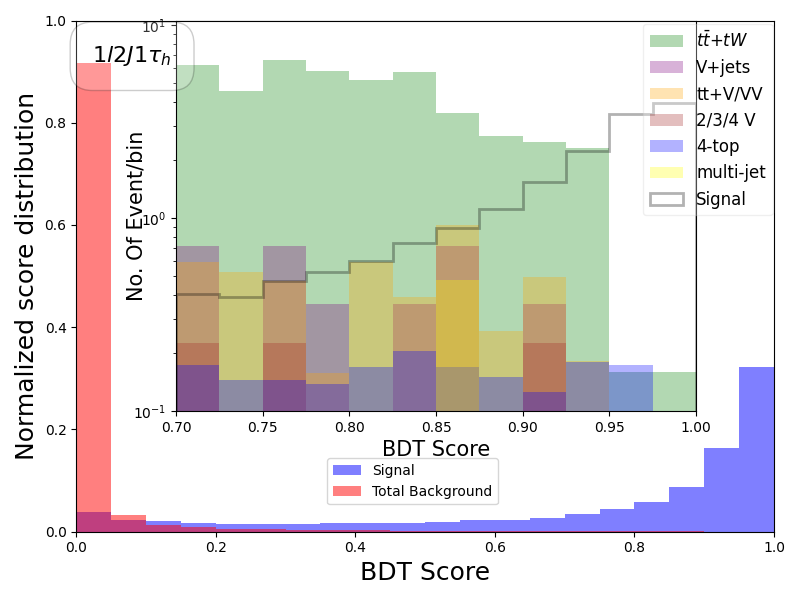}\\
    \includegraphics[width=0.32\textwidth]{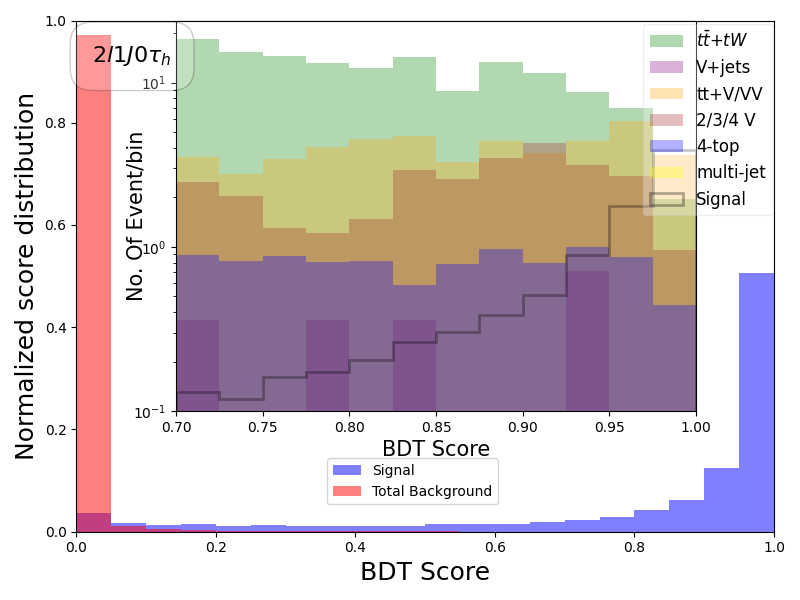}
    \includegraphics[width=0.32\textwidth]{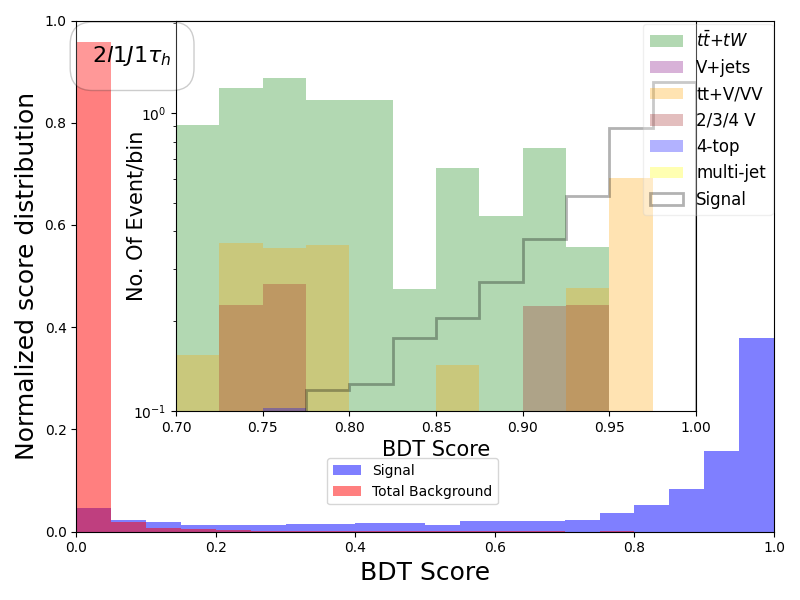}    
    \includegraphics[width=0.32\textwidth]{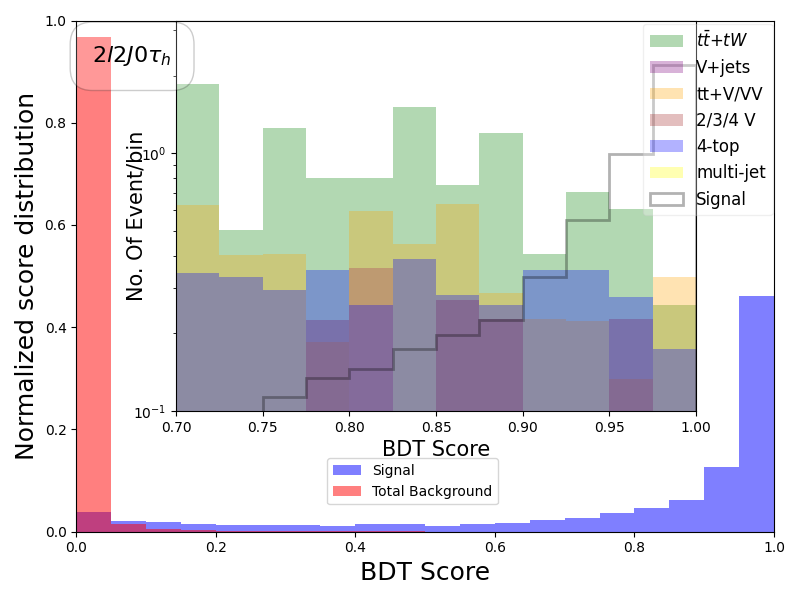}     
    \caption{The main figure shows the normalized BDT score distribution for the signal with a leptoquark mass of 1.5 TeV and the SM background for different signal regions. In the inset, we present the histogram of the score distribution in the region 0.7 to 1, divided into 12 bins, which are used for calculating the significance. The inset histogram shows the number of events from different sources of SM background and the signal at the LHC with a 14 TeV center-of-mass energy and 1000 inverse femtobarns of integrated luminosity.}
\label{fig:score_distribution}    
\end{figure}
\end{widetext}

The list of features used to train a BDT classifier is crucial for its performance and stability. As previously discussed, we have a comprehensive set of kinematic variables derived for all signal regions. The exact number of features varies by signal region. To prevent overfitting and ensure the classifier's stability across statistically independent datasets, we carefully select a subset of these features. This selection is guided by two key quantitative measures: \textit{signal-background overlap} and the \textit{correlation matrix}. The signal-background overlap evaluates a feature's ability to distinguish between signal and background and is defined as:
\[
\text{Overlap} = 1 - \frac{1}{2} \int \left| \frac{1}{\sigma_s} \frac{d\sigma_s}{dF} - \frac{1}{\sigma_b} \frac{d\sigma_b}{dF} \right| dF,
\]
where \(\sigma_s\) and \(\sigma_b\) denote the signal and background cross-sections, and \(F\) is the feature under consideration. A smaller overlap value signifies a better ability to discriminate between signal and background for that feature. The correlation matrix measures linear relationships between features to identify and eliminate redundancies. It is given by:
\[
C_{ij} = \frac{\text{Cov}(F_i, F_j)}{\Delta_{F_i} \Delta_{F_j}},
\]
where \(\text{Cov}(F_i, F_j)\) is the covariance between features \(F_i\) and \(F_j\), and \(\Delta_{F_i}\), \(\Delta_{F_j}\) are their standard deviations. Highly correlated features (\(|C_{ij}| \approx 1\)) are considered redundant, and one is typically excluded to reduce multicollinearity and overfitting. To select the optimal set of features used for training the BDT with hyperparameters detailed in Table~\ref{tab:bdt_hyperparameters}, we apply the following criteria: only features with a signal-background \(\text{Overlap} < 0.5\) and a correlation coefficient \(|C_{ij}| < 0.5\) with all other selected features are included. These thresholds ensure that the chosen features are both significantly discriminative and minimally redundant, improving the classifier's performance and robustness.

In Table~\ref{tab:feature_importance}, we present the list of the 12 most effective features for classifying signal and background events in each signal region. These features are selected based on their contribution to the BDT classifiers. The table employs color coding to indicate variable importance, which highlights the relative effectiveness of each feature in distinguishing signal from background. As expected, features such as the effective mass ($M_{eff}$) and the invariant masses ($M_{J\tau}$) of tagged boosted SM objects ($J$) and $\tau$-tagged jets are among the most effective kinematic variables for distinguishing leptoquark signals over the SM backgrounds across all signal regions. The $M_{eff}$ distribution resulting from the production of TeV-scale leptoquark pairs peaks at a significantly higher value compared to the $M_{eff}$ distribution for SM background processes. Similarly, the $M_{J\tau}$ distribution arising from the decay of leptoquarks into top-$\tau$ pairs tends to peak near the leptoquark mass, making it a robust discriminating feature. For \texttt{0 lepton} signal regions, where significant backgrounds arise from QCD multi-jet production (see Table~\ref{tab:signal_background_SR}), the top-scores ($S_T^J$) of tagged boosted SM objects are particularly effective in suppressing SM background contributions from QCD multi-jets. In contrast, for \texttt{1 lepton} or \texttt{2 lepton} signal regions, the dominant SM background arises from top-quark pair production. Since such backgrounds also include hadronically decaying boosted top quarks, $S_T^J$ is less efficient for distinguishing signals from backgrounds in these regions. Instead, various combinations of invariant masses involving $b$- or $\tau$-tagged jets and leptons play a crucial role in distinguishing signal events from background events in these signal regions. After discussing the technical details of training the BDT based multivariate classifiers to distinguish the signal events from the SM backgrounds in each signal region, we are now prepared to present the final results in the following sections.

\subsection{\label{sec:result}Result}
In Figure~\ref{fig:score_distribution}, we present the score distributions of signal and background for each BDT classifier trained for specific signal regions. These distributions are generated using statistically independent datasets for both signal and background. For the signal, the dataset corresponds to a leptoquark mass of 1.5 TeV. The testing background datasets for different background processes are generated with events equivalent to at least 1500 inverse femtobarn of luminosity. In our analysis, these BDT scores serve as the final discriminating observable, providing a quantitative measure for distinguishing between signal and background events. While the main plots in Figure~\ref{fig:score_distribution} illustrate the normalized score distributions for signal and background, demonstrating the performance of the classifiers in distinguishing signal from background, the inset histograms correspond to the actual number of expected signal and background events for an integrated luminosity of 1000 inverse femtobarn at the 14 TeV LHC.

We calculate the 95\% confidence level (CL) upper limits on the leptoquark production cross-section using the HistFactory ~\cite{Cranmer:2012sba} framework implemented with the \texttt{pyhf} library. Since this is an expected limit calculation, the observed event counts in each bin are assumed to match the predicted background counts. For each signal region, the BDT score distribution is analyzed within the range \(0.7\) to \(1.0\), as this range provides the maximum discriminating power between signal and background (see Figure~\ref{fig:score_distribution}). The score distribution is divided into 12 equally spaced bins, with each bin containing the predicted number of signal events (\(s_i\)) and background events (\(b_i\)) for a given integrated luminosity (see the insets of Figure~\ref{fig:score_distribution}).  These counts serve as inputs to \texttt{pyhf} for statistical analysis.

\begin{widetext}
\onecolumngrid

\begin{figure}[t]
    \centering
    \includegraphics[width=0.48\textwidth]{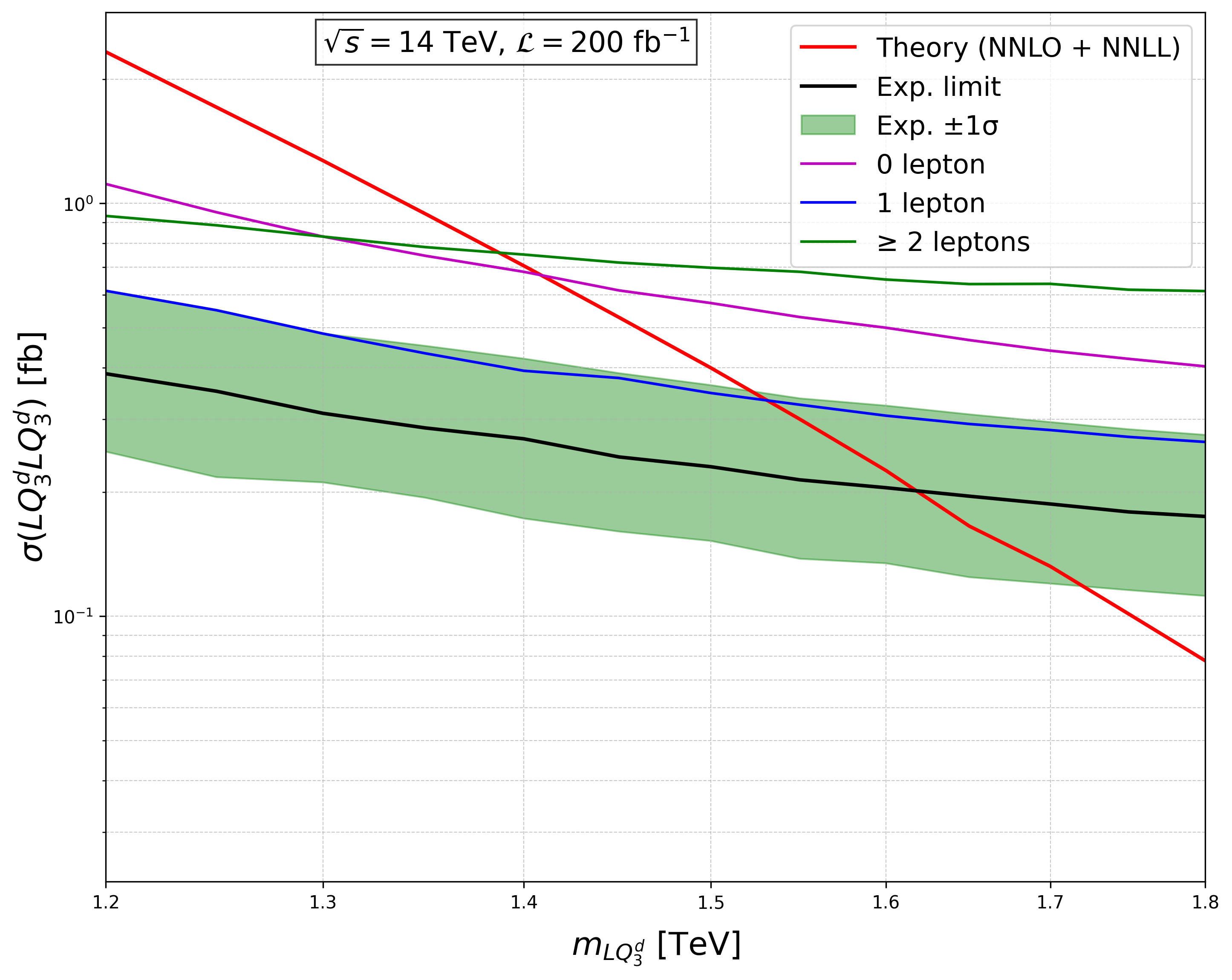}    
    \includegraphics[width=0.48\textwidth]{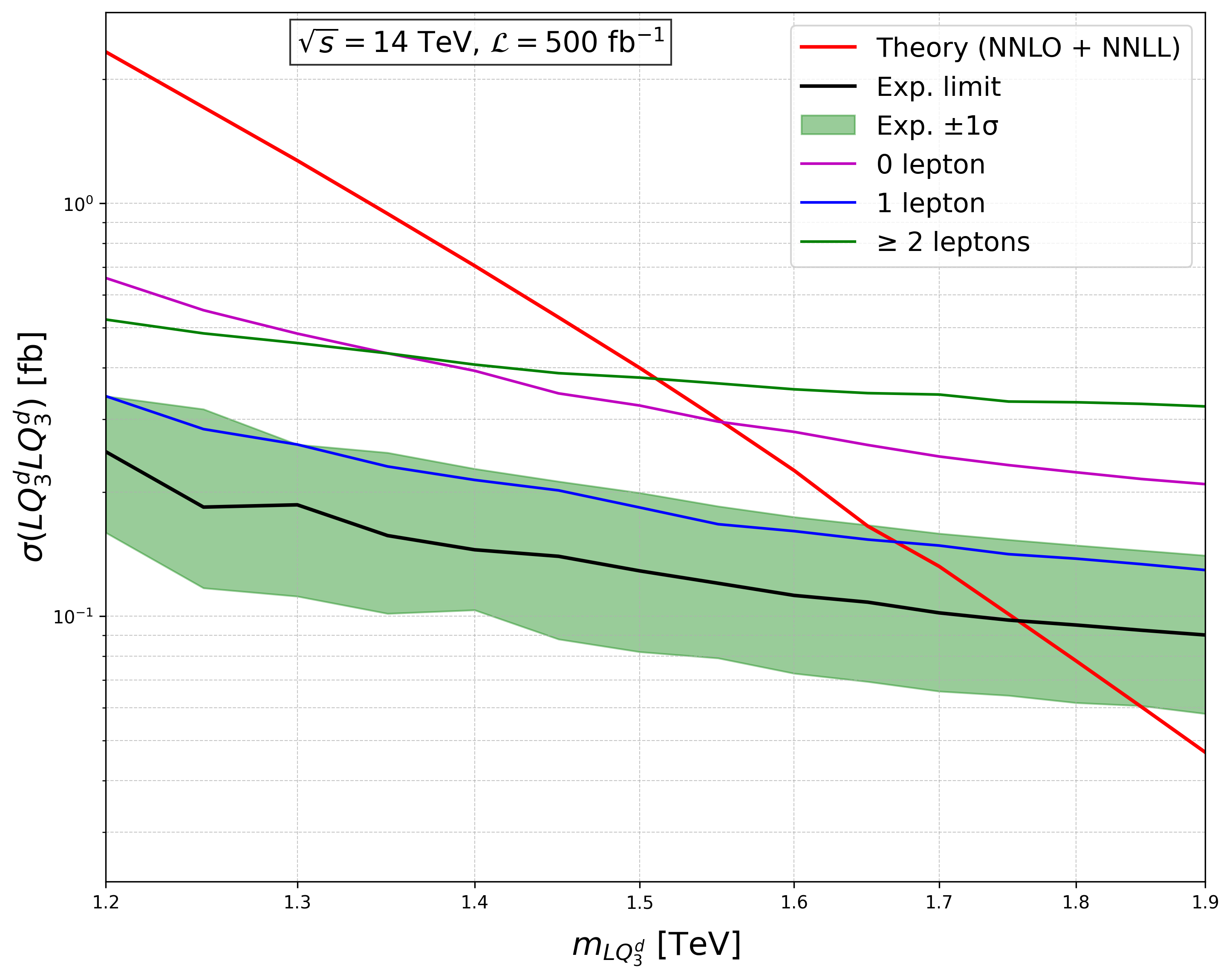}    
    \caption{Expected (dashed line) 95\% CL upper limits on the $LQ_3^d$ pair production cross-section as a function of $m_{LQ_3^d}$ resulting from the combination of all signal regions, assuming the $LQ_3^d$ decays into a top quark-$\tau$ pair with a 100\% branching ratio for 200 (left panel) and 500 (right panel) inverse femtobern luminosity of the 14 TeV LHC data. The surrounding shaded band corresponds to the $\pm 1$ standard deviation ($\pm 1 \sigma$) uncertainty around the combined expected limit. The red line shows the theoretical prediction for the leptoquark pair production cross-section at NNLO + NNLL accuracy. The individual expected limits for the 0, 1, and 2 lepton signal regions defined in Table~\ref{tab:signal_regions} are shown as the magenta, blue, and green lines, respectively.}
\label{fig:xsec_limit}    
\end{figure}
\end{widetext}

The likelihood function for a signal region is given by:
\begin{eqnarray*}
    \mathcal{L}_{\text{SR}}(\mu, \bm{\theta}) &=& \prod_{i=1}^{12} \text{Pois}\left(n_i \mid \mu s_i (1 + \theta_{s,i}) + b_i (1 + \theta_{b,i})\right) \\
    && \times \prod_{j=1}^{12} g(\theta_{s,j}, \Delta_s) g(\theta_{b,j}, \Delta_b),
\end{eqnarray*}
where \(\mu\) is the signal strength parameter scaling the signal cross-section, and \(n_i\) represents the observed (or assumed background) counts in bin \(i\). The nuisance parameters \(\theta_{s,i}\) and \(\theta_{b,i}\) account for the systematic uncertainties in signal and background estimates, modeled by the log-normal constraint: $g(\theta, \Delta) = \frac{1}{\sqrt{2\pi} \Delta} \exp\left(-\frac{\theta^2}{2\Delta^2}\right)$,
where \(\Delta_s = 0.2\) and \(\Delta_b = 0.2\) represent flat 20\% systematic uncertainties for the signal and background, respectively.

The signal strength \(\mu\) is tested using the profile likelihood ratio: $\lambda(\mu) = {\mathcal{L}(\mu, \hat{\bm{\theta}}_{\mu})}/{\mathcal{L}(\hat{\mu}, \hat{\bm{\theta}})}$,
where \(\hat{\bm{\theta}}_{\mu}\) are the nuisance parameters that maximize the likelihood for a fixed \(\mu\), and \(\hat{\mu}\) and \(\hat{\bm{\theta}}\) are the values that globally maximize the likelihood. The test statistic is defined as:
\[
q_{\mu} =
\begin{cases}
-2 \ln \lambda(\mu) & \text{if } \hat{\mu} \leq \mu, \\
0 & \text{if } \hat{\mu} > \mu.
\end{cases}
\]
The 95\% CL upper limit on the signal strength \(\mu\) is determined using the CL\(_s\) method, defined as:
\[
\text{CL}_s = \frac{\text{CL}_{s+b}}{\text{CL}_b},
\]
where \(\text{CL}_{s+b}\) and \(\text{CL}_b\) represent the confidence levels for the signal+background and background-only hypotheses, respectively. The value of \(\mu\) is varied until \(\text{CL}_s = 0.05\), corresponding to the 95\% CL upper limit. The likelihoods for all signal regions are combined to maximize sensitivity, as the signal regions are defined to be mutually exclusive and statistically independent. The combined likelihood is given by:
\[
\mathcal{L}_{\text{combined}}(\mu, \bm{\theta}) = \prod_{\text{SR}} \mathcal{L}_{\text{SR}}(\mu, \bm{\theta}_{\text{SR}}).
\]


The final 95\% CL expected upper limits on the leptoquark production cross-section as a function of leptoquark mass are shown in Figure~\ref{fig:xsec_limit} for two luminosity scenarios: 200 and 500 \(\text{fb}^{-1}\) of 14 TeV LHC data. These limits result from the combination of all signal regions, assuming that the leptoquark decays exclusively into a top quark and a \(\tau\)-lepton. The individual expected limits for the \texttt{0 lepton}, \texttt{1 lepton}, and \texttt{2 lepton} signal regions, as well as the theoretical cross-section prediction at NNLO+NNLL accuracy, are also shown. The surrounding shaded band represents the \(\pm 1 \sigma\) uncertainty around the combined expected limit.  Figure~\ref{fig:xsec_limit} shows a significant improvement in LHC sensitivity for leptoquark searches, with the expected 95\% CL reach of the \( LQ_d^3 \) mass approaching up to 1.63 (1.77) TeV at the LHC with a center of mass energy of 14 TeV and integrated luminosity of 200 (500) fb\(^{-1}\).

\section{\label{sec:conclusion}Summary and Outlook}
In this work, we explored the potential of a machine learning-based approach for boosted object tagging and event selection in carefully designed signal regions, categorized based on the multiplicity of tagged boosted SM jets, $\tau$-jets, and leptons, to enhance the discovery prospects of third-generation scalar leptoquarks (\(LQ_3^d\)) at the 14 TeV LHC. By focusing on their decays into top quarks and tau leptons with a 100\% branching fraction, we defined comprehensive signal regions tailored to capture a wide range of final-state topologies. A combination of carefully constructed universal and signal region-specific kinematic features was employed to train multivariate classifiers for each signal region. We observe a significant improvement in the LHC sensitivity for leptoquark searches, with the expected 95\% CL reach of the \( LQ_d^3 \) mass approaching up to 1.63 (1.77) TeV at the LHC with a center-of-mass energy of 14 TeV and integrated luminosity of 200 (500) fb\(^{-1}\).

\bibliography{apssamp}

\end{document}